\documentclass[iop, apj, twocolappendix, numberedappendix, appendixfloats]{emulateapj}
\usepackage{threeparttable}
\usepackage{multirow}
\usepackage{times}
\usepackage{enumitem}
\usepackage{url} 
\usepackage{amsmath,bm}

\shorttitle{}
\shortauthors{Y. Huang et al.}

\begin{document}

\title{Beyond spectroscopy. II. Stellar parameters for over twenty million stars in the northern sky from SAGES\,DR1 and Gaia\,DR3}
\author{Yang Huang\altaffilmark{1,2}}
\author{Timothy C. Beers\altaffilmark{3}}
\author{Haibo Yuan\altaffilmark{4}}
\author{Ke-Feng Tan\altaffilmark{2}}
\author{Wei Wang\altaffilmark{2}}
\author{Jie Zheng\altaffilmark{2}}
\author{Chun Li\altaffilmark{2}}
\author{Young Sun Lee\altaffilmark{5}}
\author{Hai-Ning Li\altaffilmark{2}}
\author{Jing-Kun Zhao\altaffilmark{2}}
\author{Xiang-Xiang Xue\altaffilmark{2}}
\author{Yujuan Liu\altaffilmark{2}}
\author{Huawei Zhang\altaffilmark{6,7}}
\author{Xue-Ang Sun\altaffilmark{8}}
\author{Ji Li\altaffilmark{8}}
\author{Hong-Rui Gu\altaffilmark{2,1}}
\author{Christian Wolf\altaffilmark{9,10}}
\author{Christopher A. Onken\altaffilmark{9,10}}
\author{Jifeng Liu\altaffilmark{2,1}}
\author{Zhou Fan\altaffilmark{2,1}}
\author{Gang Zhao\altaffilmark{2,1}}

\altaffiltext{1}{School of Astronomy and Space Science, University of Chinese Academy of Sciences, Beijing 100049, Chinese; hunagyang@ucas.ac.cn}
\altaffiltext{2}{Key Lab of Optical Astronomy, National Astronomical Observatories, Chinese Academy of Sciences, Beijing 100012, P.\,R.\,China;  zfan@bao.ac.cn; gzhao@nao.ac.cn}
\altaffiltext{3}{Department of Physics and Astronomy and JINA Center for the Evolution of the Elements (JINA-CEE), University of Notre Dame, Notre Dame, IN 46556, USA}
\altaffiltext{4}{Department of Astronomy, Beijing Normal University, Beijing 100875, People's Republic of China}
\altaffiltext{5}{Department of Astronomy and Space Science, Chungnam National University, Daejeon 34134, Republic of Korea}
\altaffiltext{6}{Department of Astronomy, School of Physics, Peking University, Beijing 100871, People's Republic of China}
\altaffiltext{7}{Kavli Institute for Astronomy and Astrophysics, Peking University, Beijing 100871, People's Republic of China}
\altaffiltext{8}{Department of Space Science and Astronomy, Hebei Normal University, Shijiazhuang 050024, People's Republic of China}
\altaffiltext{9}{Research School of Astronomy and Astrophysics, Australian National University, Canberra, ACT 2611, Australia}
\altaffiltext{10}{Centre for Gravitational Astrophysics, Research Schools of Physics, and Astronomy and Astrophysics, Australian National University, Canberra, ACT 2611, Australia}

\begin{abstract}
We present precise photometric estimates of stellar parameters, including effective temperature, metallicity, luminosity classification, distance, and stellar age, for nearly 26 million stars using the methodology developed in the first paper of this series, based on the stellar colors from the Stellar Abundances and Galactic Evolution Survey (SAGES) DR1 and {\it Gaia} EDR3. 
The optimal design of stellar-parameter sensitive $uv$ filters by SAGES has enabled us to determine photometric-metallicity estimates down to $-3.5$, similar to our previous results with the SkyMapper Southern Survey (SMSS), yielding a large sample of over five million metal-poor (MP; [Fe/H]\,$\le -1.0$) stars and nearly one million very metal-poor (VMP; [Fe/H]\,$\le -2.0$) stars.
The typical precision is around $0.1$\,dex for both dwarf and giant stars with [Fe/H]\,$>-1.0$, and 0.15-0.25/0.3-0.4\,dex for dwarf/giant stars with [Fe/H]\,$<-1.0$.
Using the precise parallax measurements and stellar colors from {\it Gaia}, effective temperature, luminosity classification, distance and stellar age are further derived for our sample stars.
This huge data set in the Northern sky from SAGES, together with similar data in the Southern sky from SMSS, will greatly advance our understanding of the Milky Way, in particular its formation and evolution.   
\end{abstract}
\keywords{Galaxy: stellar content -- Galaxy: halo -- stars: fundamental parameters -- stars: distances -- stars:abundances -- methods: data analysis}

\section{Introduction}
Estimates of stellar parameters, in particular the metallicity, of a large, complete sample of stars is of vital importance to understand the formation and evolution of the Milky Way.
In the past decades, massive progress has been achieved by large-scale spectroscopic surveys, such as the HK Survey \citep[]{1985AJ.....90.2089B, 1992AJ....103.1987B}, the Hamburg/ESO Survey (HES; \citealt{2003RvMA...16..191C}) the Sloan Digital Sky Survey (SDSS; \citealt{2000AJ....120.1579Y}), the Radial Velocity Experiment (RAVE; \citealt{2006AJ....132.1645S}), the Large Sky Area Multi-Object Fiber Spectroscopic Telescope (LAMOST; \citealt[]{2012RAA....12..735D, 2014IAUS..298..310L}), the Galactic Archaeology with HERMES project (GALAH; \citealt{2015MNRAS.449.2604D}), and the Apache Point Observatory Galactic Evolution Experiment (APOGEE; \citealt{2017AJ....154...94M}).
However, the total number of observed targets collected from all those surveys is no greater than about ten million, less than one ten-thousandth of the estimated total numbers of Milky Way stars.
This under-sampling, together with the complex target-selection strategies, makes it extremely difficult to understand the full assembly history of our Galaxy.  

In the first paper of this series \citep[][hereafter H22]{2022ApJ...925..164H}, we proposed to alleviate this issue of current spectroscopic surveys by deriving stellar parameters for a huge number of stars using narrow/medium-bandwidth photometric surveys (see Table\,1 of H22 for a summary).
As a pioneering experiment, H22 present measurements of stellar parameters, including metallicity, luminosity classification, effective temperature, 
distance, and stellar age, for over 24 million stars, based on the stellar colors from the SkyMapper Southern Survey (SMSS; \citealt[]{2018PASA...35...10W, 2019PASA...36...33O}) and {\it Gaia} \citep{2021A&A...649A...1G}, as well as the parallax measurements from {\it Gaia}.
This huge data set has already been applied to a number of Galactic studies, including searching for metal-poor stars \citep{2022ApJ...927...13Z}, discovery of ancient halo substructures \citep{2022ApJS..261...19S, 2022ApJ...926...26S, 2022ApJ...930..103Y}, and understanding the disk/halo formation history (Hong et al. 2023). Its contribution to this field is just beginning to be explored.

In this paper, we present a second pioneering experiment in the Northern sky, using the data from the first data release of the Stellar Abundance and Galactic Evolution Survey \citep[SAGES DR1;][]{2023arXiv230615611F} and {\it Gaia} EDR3  \citep{2021A&A...649A...1G}.
SAGES is an optical multi-band ($u, v, g, r, i, {\rm DDO}\text{-51}$, H$\alpha_{\rm wide}$, H$\alpha_{\rm narrow}$) large-scale  photometric survey, aiming to cover 12,000 square degrees of the Northern sky with $\delta > -5^{\circ}$ down to a  $5\sigma$ depth of 21.5 in the $u$-band \citep{2018RAA....18..147Z}.
The $u$-band filter is the same as in the Str{\"o}mgren system \citep{1956VA......2.1336S}, and the $v$-band is optimized to provide reliable metallicity measurements by shifting the central wavelength of the SkyMapper $v$ \citep{2011PASP..123..789B} to longer wavelengths, by about 100\,\AA, to reduce the effect of molecular bands of carbon and nitrogen on the metallicity estimates.

The special design of the $uv$ filters (especially the $v$-band) provides photometric sensitivity to stellar surface gravity and metallicity that are well-demonstrated by numerous previous efforts with similar filter systems (e.g., \citealt{2004A&A...418..989N,  2017MNRAS.471.2587S, 2019MNRAS.482.2770C, 2019ApJS..243....7H, 2021ApJS..254...31C}; H22).
The $gri$ filters are SDSS-like, which can be used to estimate the stellar effective temperature.
The combination of H$\alpha$ and other filters can be used to estimate the values of reddening. 
Similar to our effort with SMSS (H22), here we present stellar parameter estimates for over 26 million stars using the $uv$-band data released in SAGES DR1, along with the photometric and parallax information provided by {\it Gaia} EDR3  \citep[]{2021A&A...649A...1G}.

This paper is structured as follows. 
In Section\,2, we introduce the data adopted in the current work.
In Section\,3, photometric-metallicity estimates from the stellar colors of SAGES DR1 and {\it Gaia} EDR3 are described, along with various checks on the photometric measurements.
The determinations of effective temperature, $T_{\rm eff}$, distance, and age are presented in Section\,4.
Radial velocity measurements collected from previous spectroscopic surveys and the final sample are described in Section\,5.
We present a summary in  Section\,6.

\begin{figure*}
\begin{center}
\includegraphics[scale=0.35,angle=0]{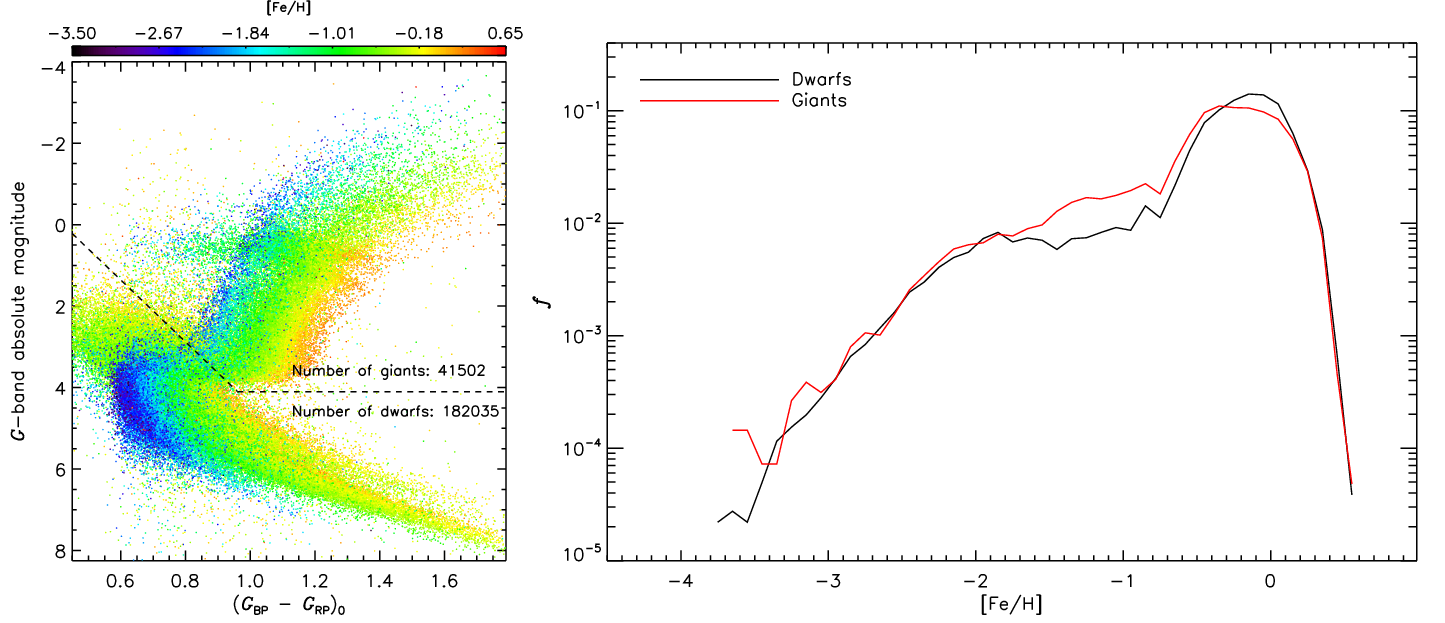}
\caption{{\it Left panel:} Hertzsprung–Russell (H-R) diagram, $M_{G_0}$ versus $(G_{\rm BP} - G_{\rm RP})_0$, of the stars in the training set defined in Section\,3.1,  color-coded by [Fe/H] as shown in the color bar at the top.
The dashed lines represent the empirical cuts $M_{G_0} = -3.20 + 7.60\cdot (G_{\rm BP} - G_{\rm RP})_0$ or $M_{G_0} = 4.1$, used to separate dwarf and giant stars.
{\it Right panel:} Metallicity ([Fe/H]) distributions of dwarf (black line) and giant (red line) stars in the training sample.}
\end{center}
\end{figure*}

\begin{figure*}
\begin{center}
\includegraphics[scale=0.40,angle=0]{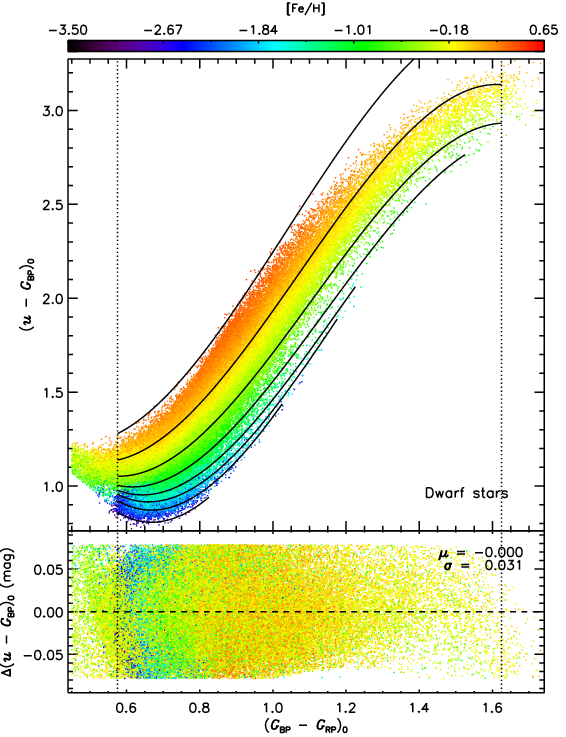}
\includegraphics[scale=0.40,angle=0]{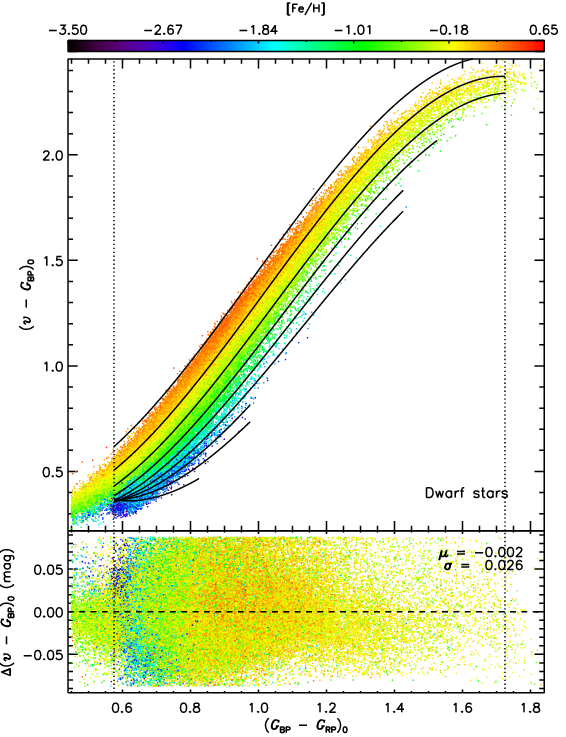}
\includegraphics[scale=0.40,angle=0]{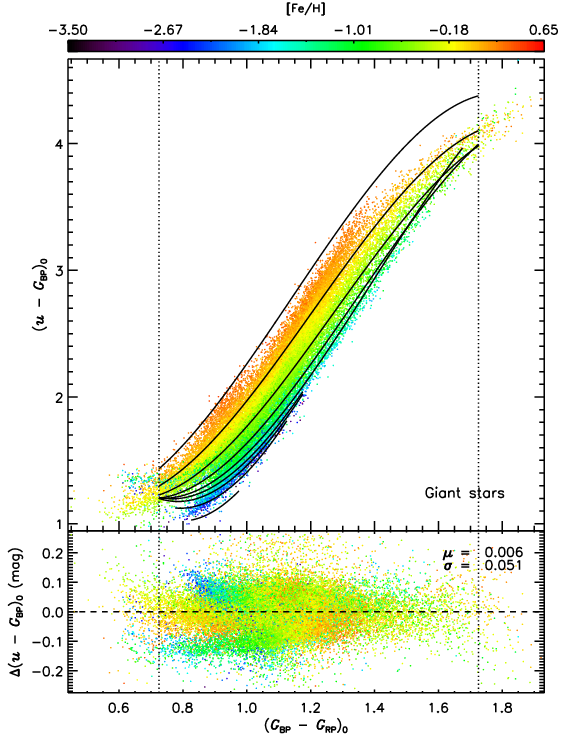}
\includegraphics[scale=0.40,angle=0]{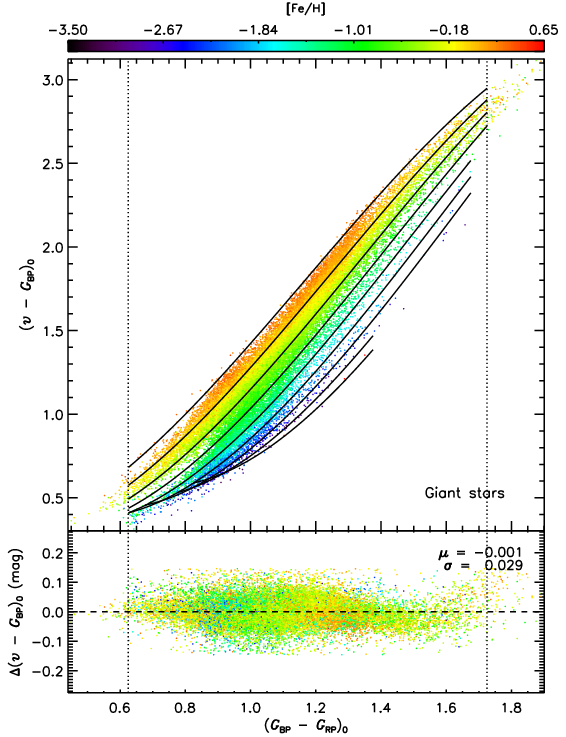}
\caption{\textbf{Top:} Metallicity-dependent stellar loci of training-set dwarf stars in the plane of $(u - G_{\rm BP})_0$ versus $(G_{\rm BP} - G_{\rm RP})_0$ (left panel) and  $(v - G_{\rm BP})_0$ versus $(G_{\rm BP} - G_{\rm RP})_0$ (right panel), color-coded by [Fe/H] as shown in the top color bars.
The black lines represent our best fits for [Fe/H] with values ranging from $+0.5$ (top) to  $-3.5$ (bottom) in steps of $0.5$\,dex, as described by Equation\,1.
The dashed lines mark the color region in $(G_{\rm BP} - G_{\rm RP})_0$ for which the data points yield robust fits.
The lower part of each panel shows the fit residual, as a function of color $(G_{\rm BP} - G_{\rm RP})_0$, with the values of median and standard deviation of the residual marked in the top-right corner.
\textbf{Bottom:} Similar to the top panels, but for training-set giant stars.}
\end{center}
\end{figure*}

\begin{figure*}
\begin{center}
\includegraphics[scale=0.335,angle=0]{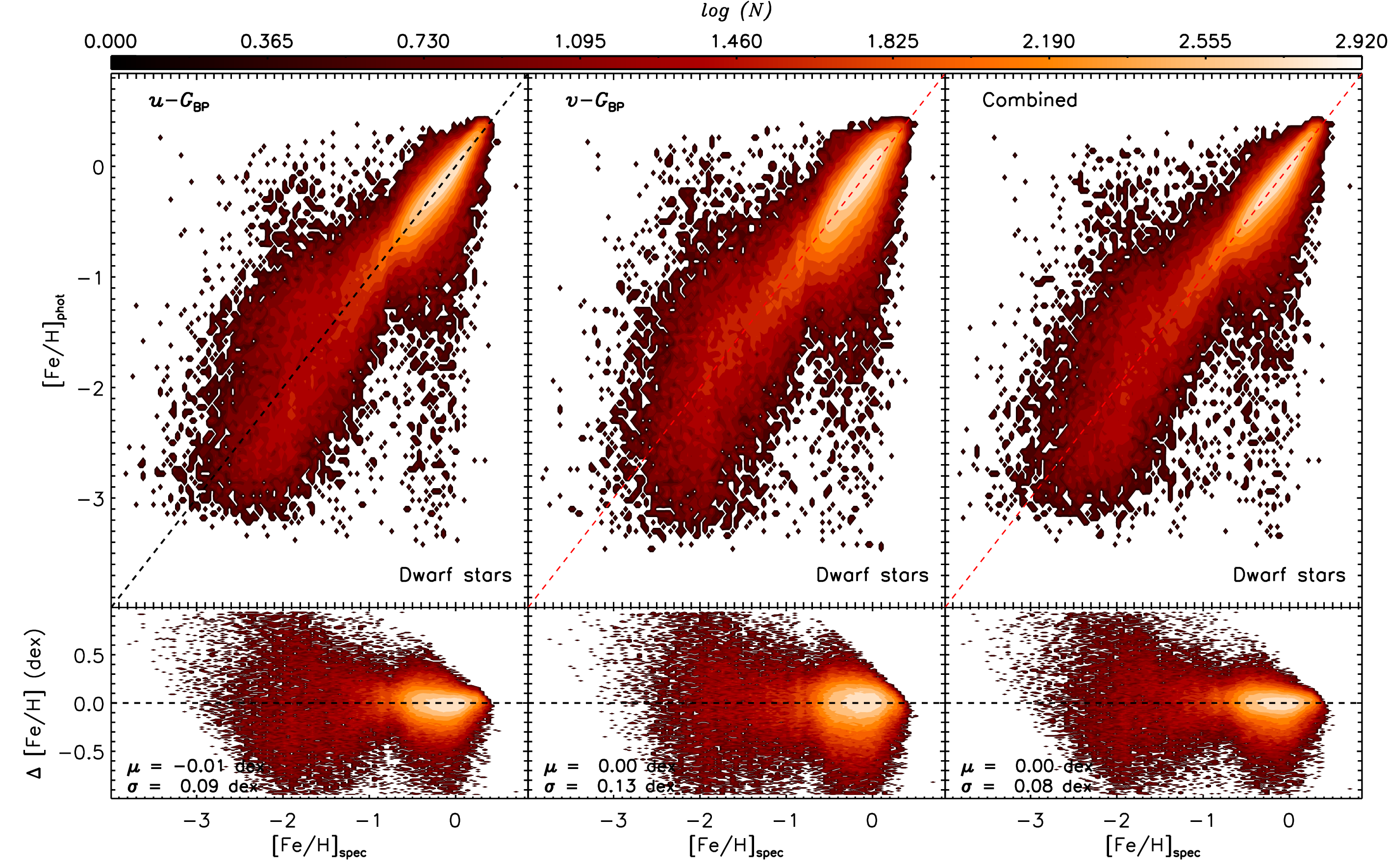}
\includegraphics[scale=0.335,angle=0]{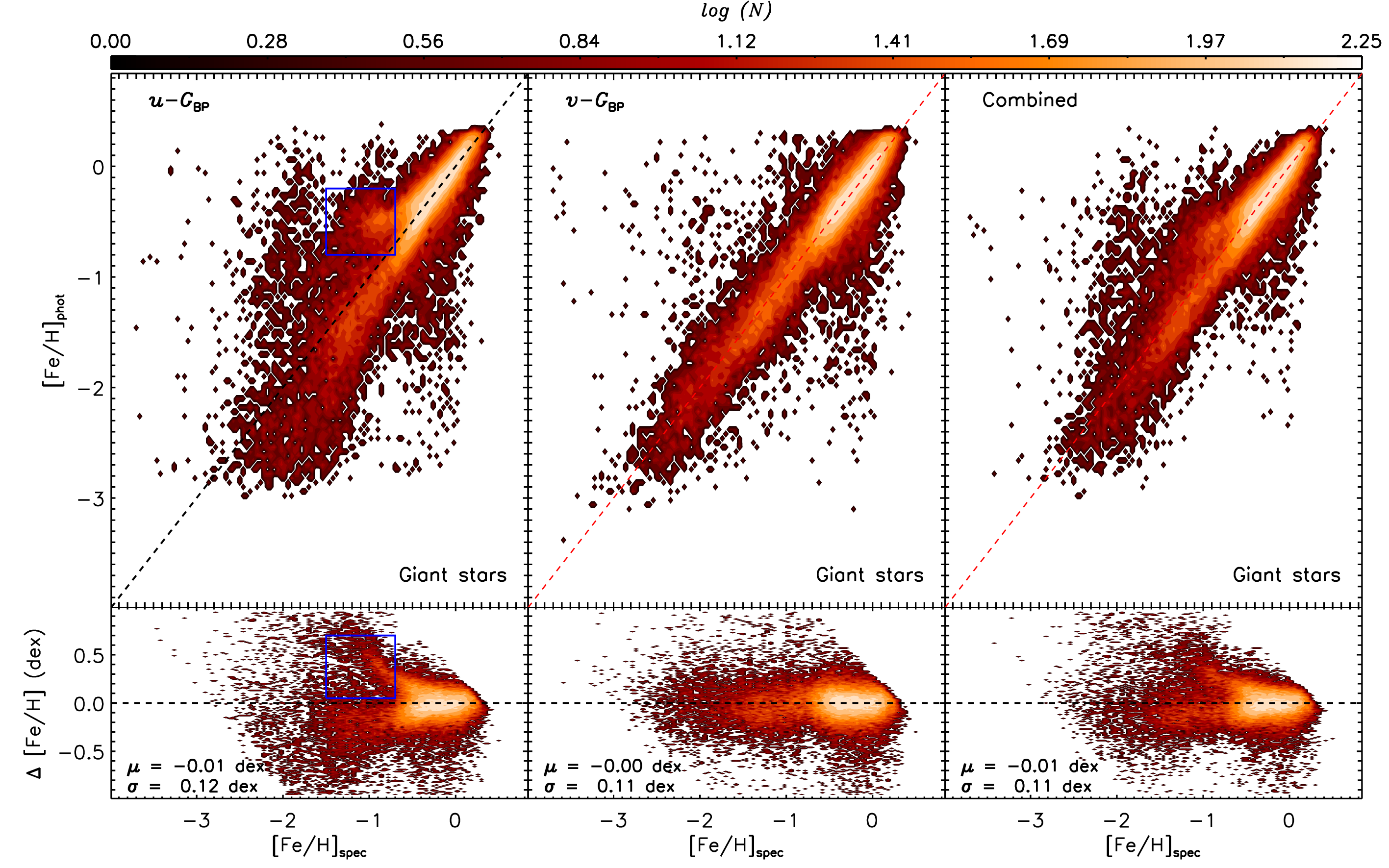}
\caption{\textbf{Top:} Internal check between spectroscopic metallicity and the photometric metallicity derived from the colors $u - G_{\rm BP}$ (left panel), $v - G_{\rm BP}$ (middle panel), and the combination of the two colors (right panel), using a maximum-likelihood approach (see Section\,3.2) for the training-set dwarf stars.
The metallicity difference (photometric minus spectroscopic), as a function of the spectroscopic metallicity, is shown in the lower part of each panel, with the median and standard deviation of the difference marked in the bottom-left corner.
In each panel, a color-coded contour of the stellar number density on a logarithmic scale is shown.
\textbf{Bottom:} Similar to the top panels, but for the training-set giant stars. 
In the left panel, a region of stars marked by the blue box with large deviations from the spectroscopic estimates are mainly from warm, low-gravity blue giants, e.g., the blue horizontal-branch (BHB) stars.}
\end{center}
\end{figure*}

\begin{figure*}
\begin{center}
\includegraphics[scale=0.415,angle=0]{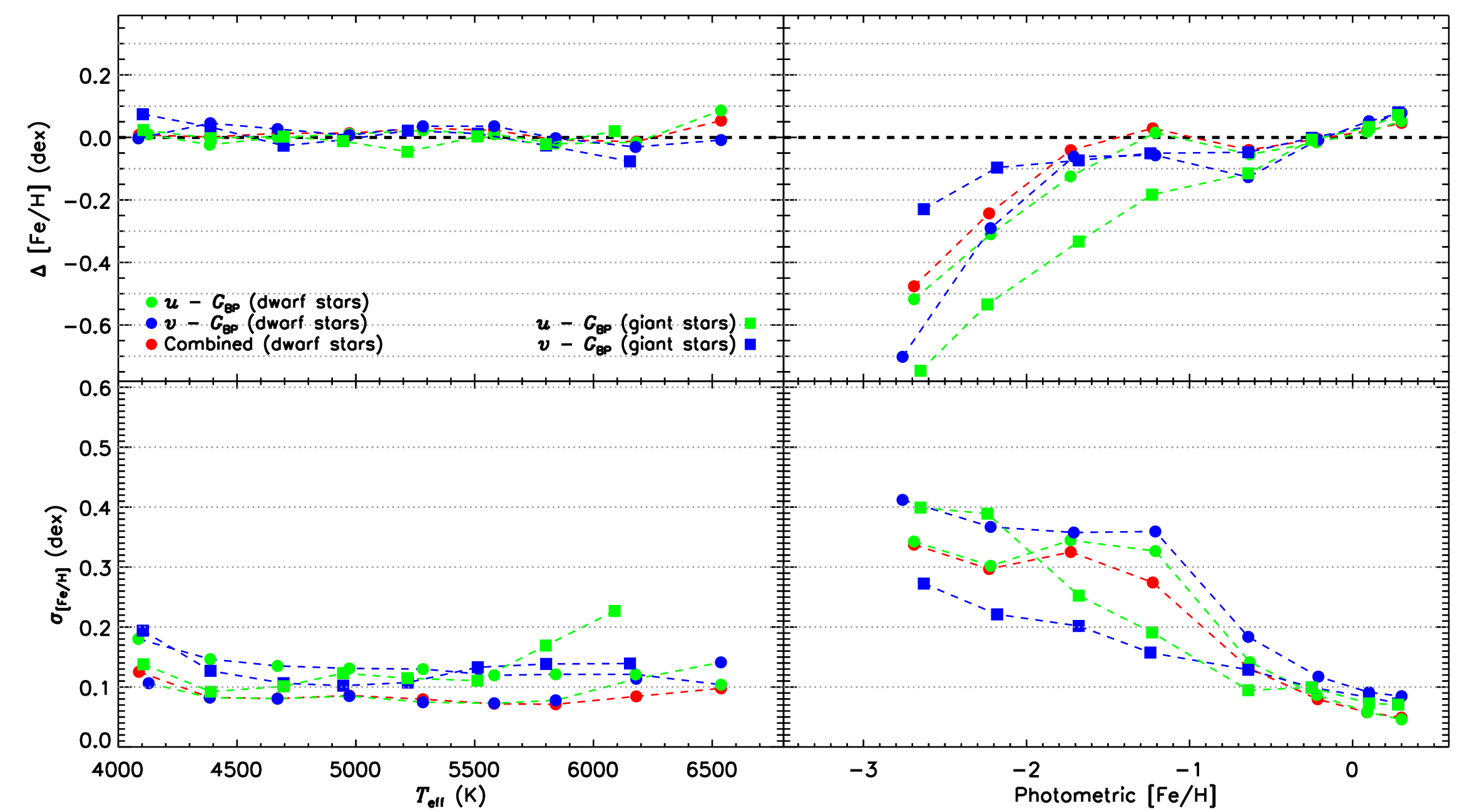}
\caption{Median offsets (top panels) and standard deviations (bottom panels) of the metallicity differences (photometric minus spectroscopic), as a function of effective temperature (left panels) and photometric [Fe/H] (right panels), as calculated from the training sample (dots for dwarf stars and squares for giant stars). Different symbol colors indicate the metallicity determined from different stellar colors.}
\end{center}
\end{figure*}

\begin{figure}
\begin{center}
\includegraphics[scale=0.25,angle=0]{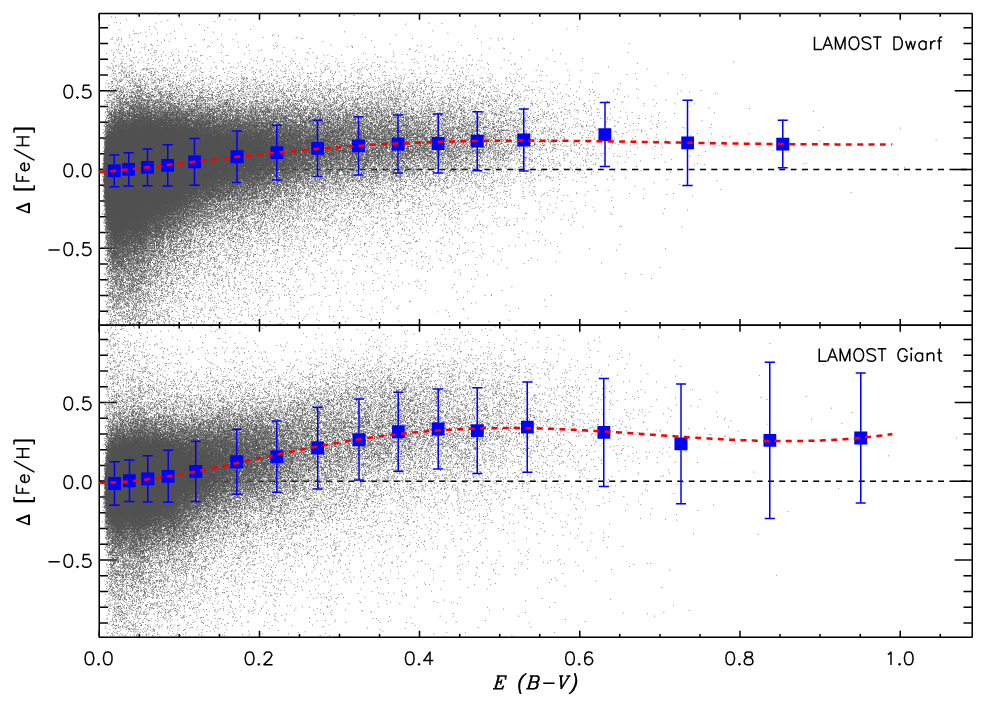}
\caption{Metallicity difference (photometric minus spectroscopic), as a function of SFD $E (B-V)$, for over 600 thousand and 200 thousand LAMOST dwarf (top) and giant (bottom) stars, respectively.
The blue squares and error bars in each panel represent the median and dispersion of the metallicity differences in the individual $E (B-V)$ bins.
The red dashed lines in each panel represent the best-fit fifth-order polynomials to the trends of metallicity differences with $E (B-V)$.}
\end{center}
\end{figure}

\begin{figure*}
\begin{center}
\includegraphics[scale=0.355,angle=0]{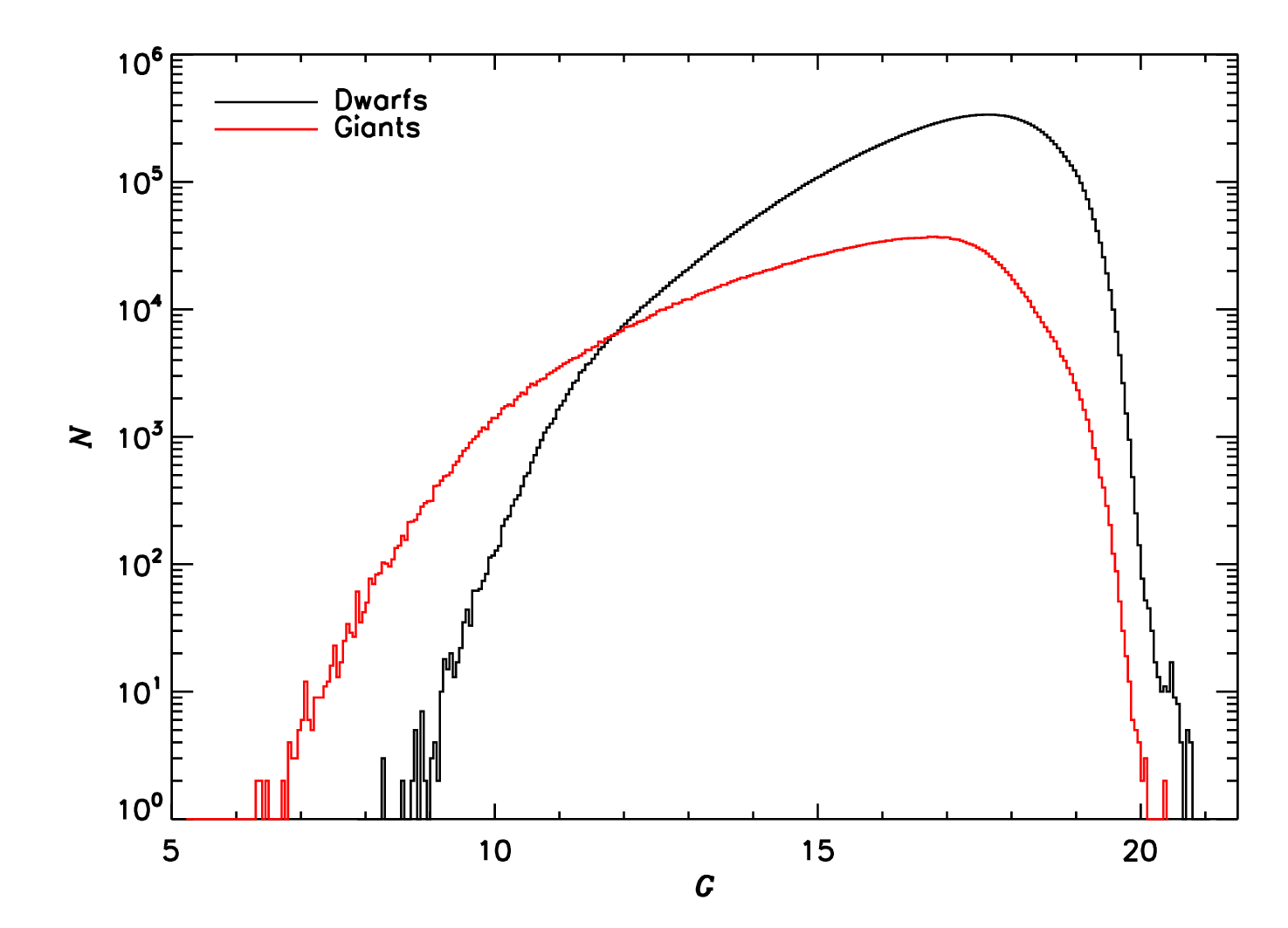}
\includegraphics[scale=0.355,angle=0]{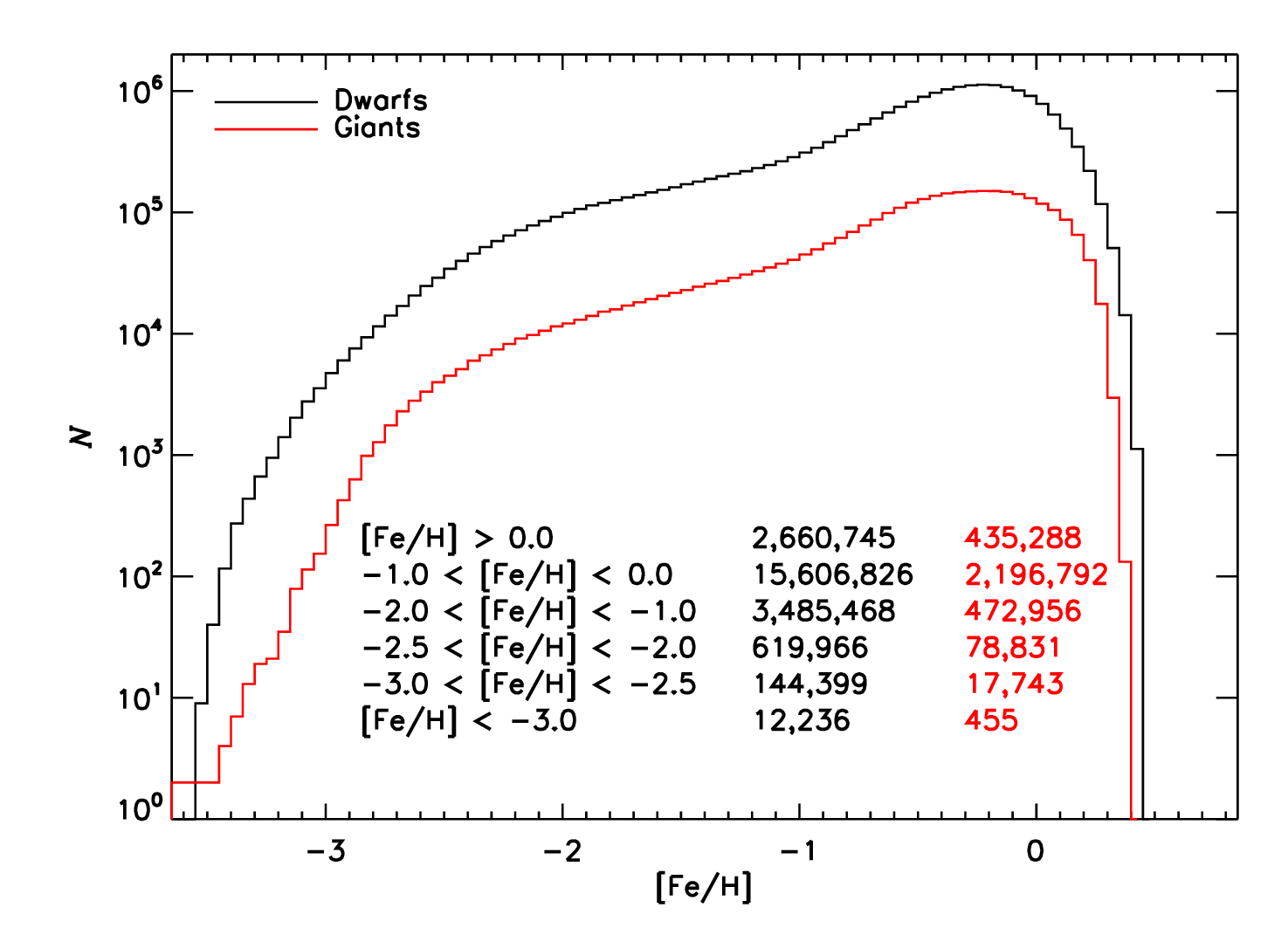}
\caption{{\it Left panel:} Magnitude distributions of dwarf (black line) and giant (red line) stars with photometric metallicity estimated from the colors provided by SAGES DR1 and {\it Gaia} EDR3.
{\it Right panel:} Distributions of photometric metallicity for dwarf (black line) and giant (red line) stars. 
The number of dwarf (black) and giant (red) stars included in individual metallicity bins are marked.}
\end{center}
\end{figure*}

\begin{figure}
\begin{center}
\includegraphics[scale=0.30,angle=0]{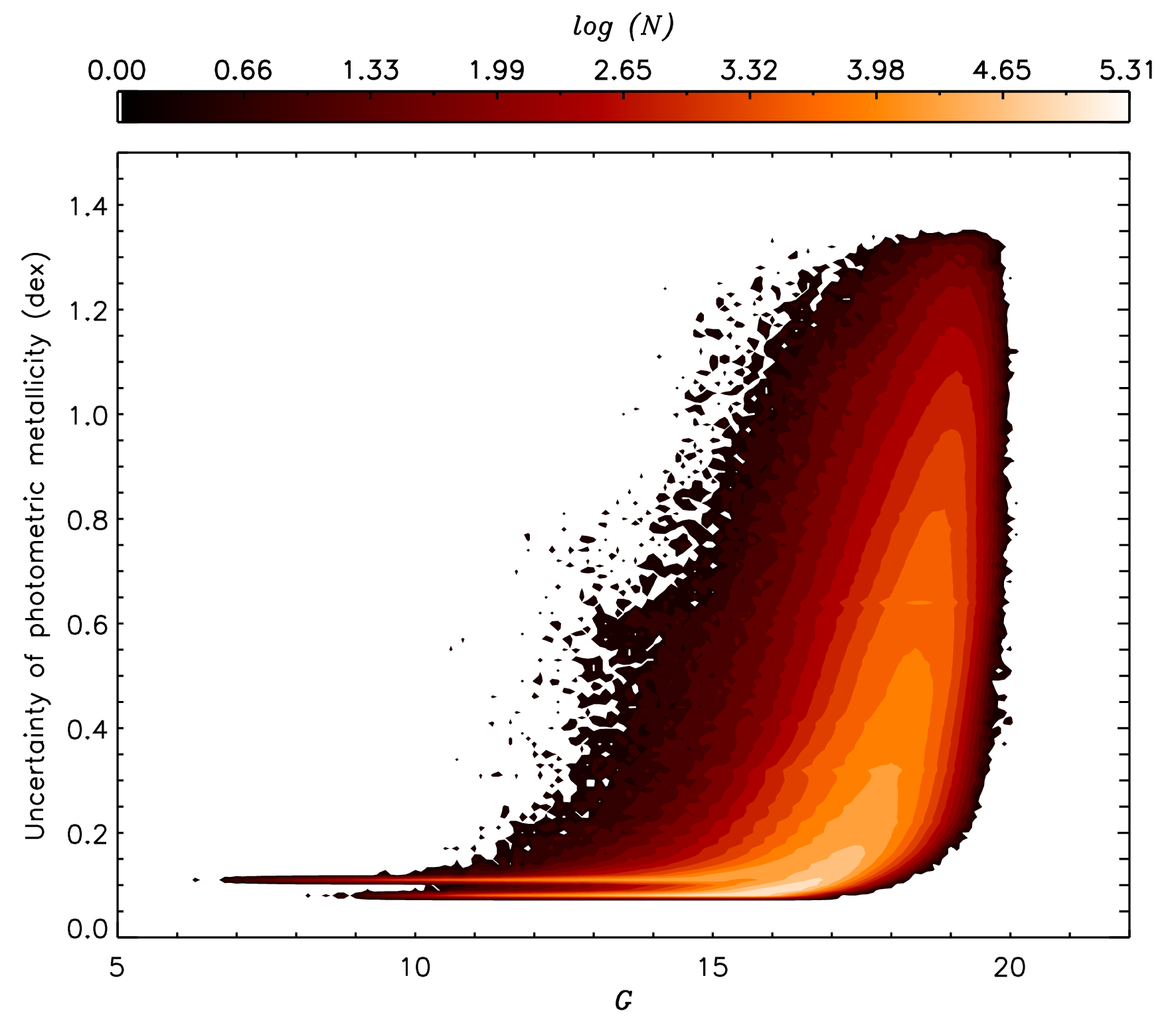}
\caption{Density map (on a logarithmic scale) of the uncertainties of 
photometric-metallicity estimates versus $G$-band magnitude.}
\end{center}
\end{figure}

\begin{figure*}
\begin{center}
\includegraphics[scale=0.295,angle=0]{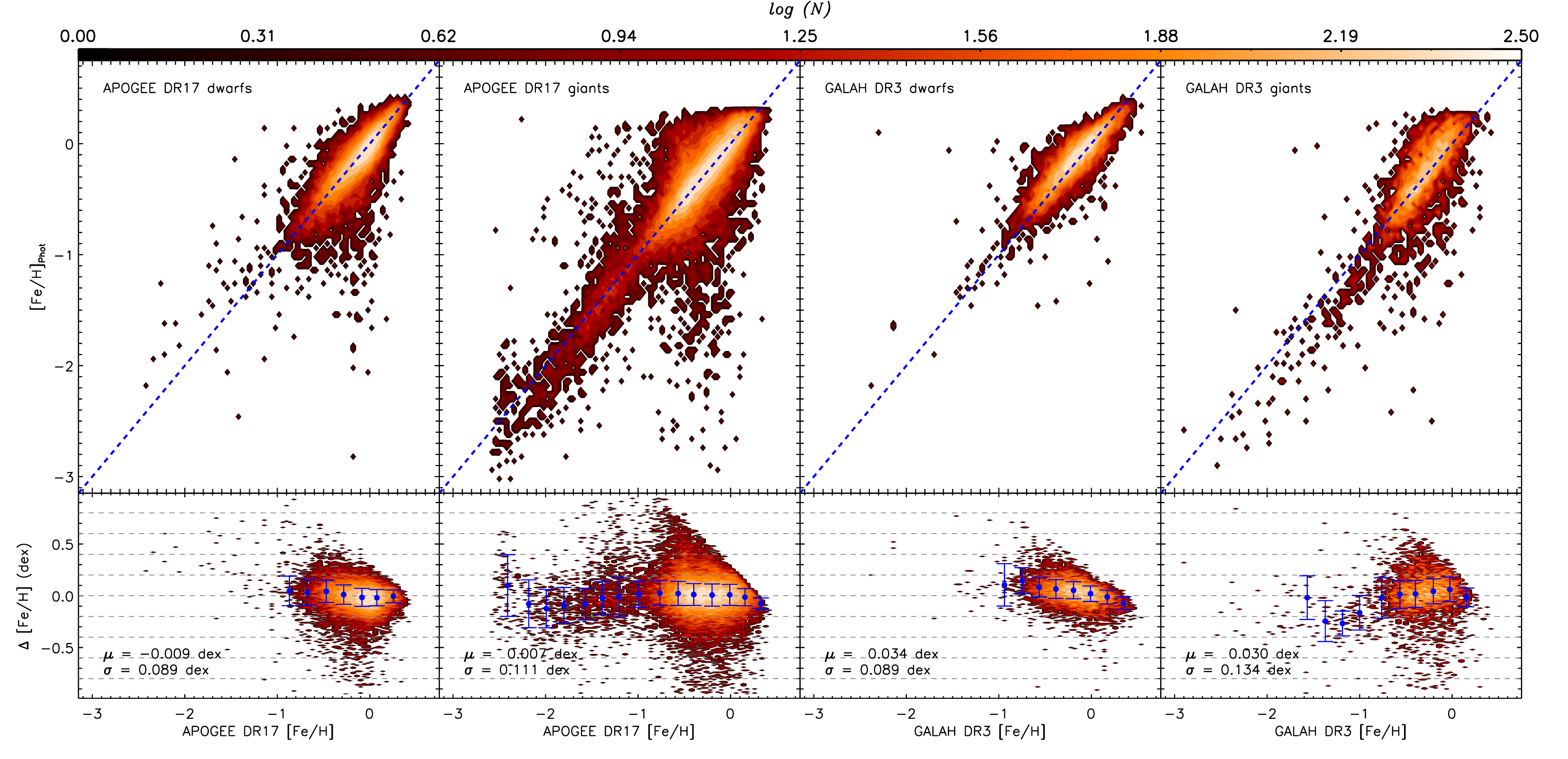}
\caption{Comparisons of photometric-metallicity estimates with high-resolution spectroscopic metallicities from APOGEE DR17 (left two panels) and GALAH DR3+ (right two panels).
The metallicity differences (photometric minus spectroscopic) are shown in the lower panels, with the overall median and standard deviation marked in the bottom-left corners.
In each panel, a color-coded contour of the stellar number density on a logarithmic scale is shown.
The blue dots and error bars in each panel indicate the medians and dispersions of the metallicity differences in the individual metallicity bins.}
\end{center}
\end{figure*}

\begin{figure}
\begin{center}
\includegraphics[scale=0.25,angle=0]{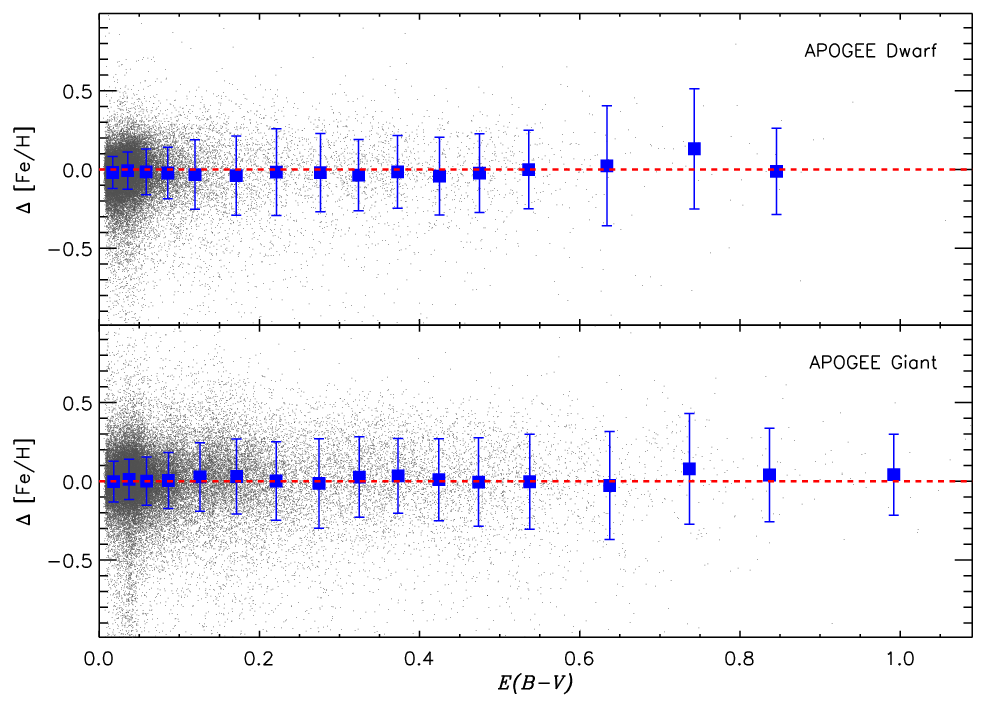}
\caption{Similar to Fig.\,7, but for APOGEE stars. Here, the $E (B-V)$ dependent offsets, as found in Fig.\,7, have been corrected to obtain the 
photometric-metallicity estimates.}
\end{center}
\end{figure}

\begin{figure}
\begin{center}
\includegraphics[scale=0.35,angle=0]{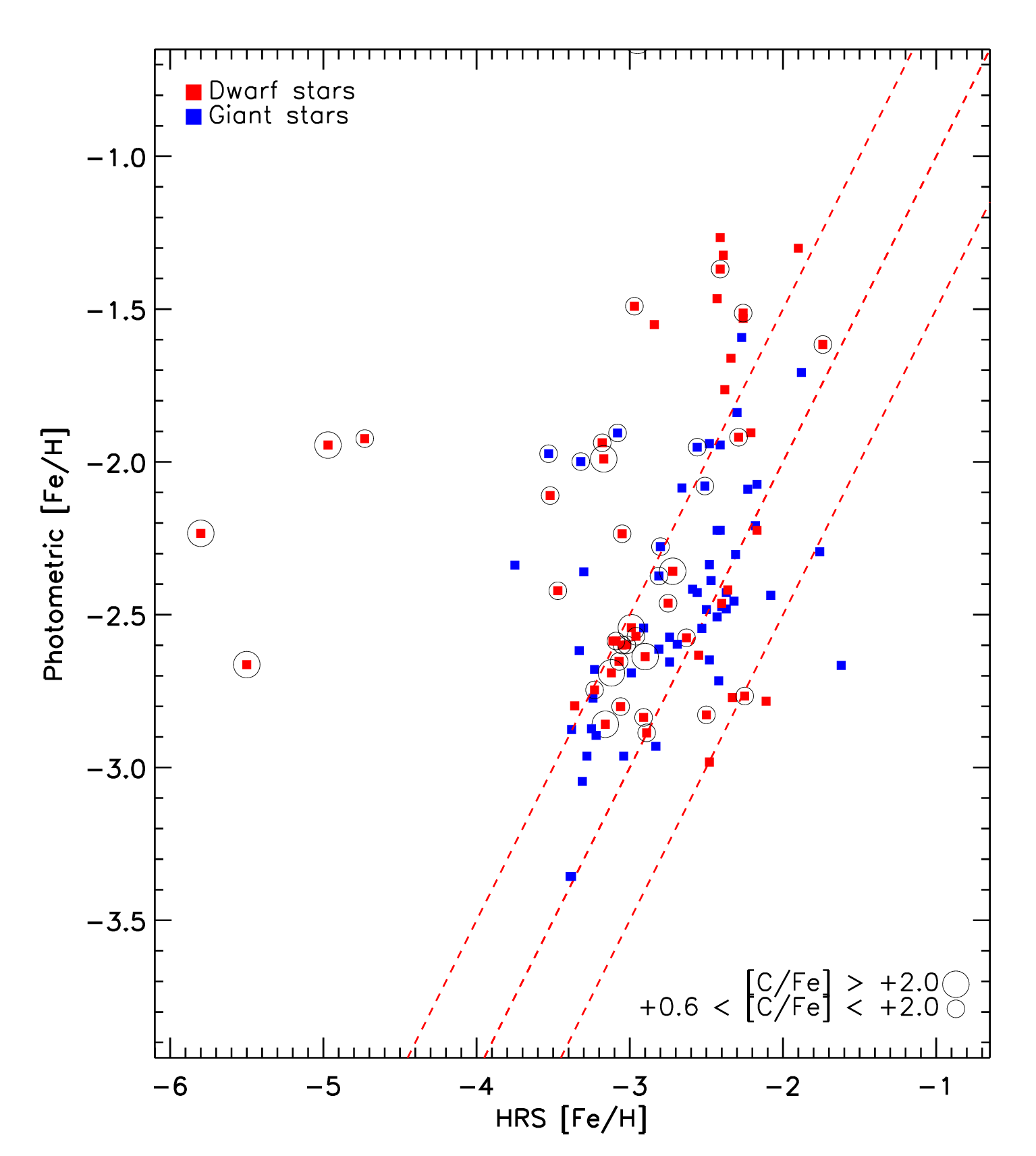}
\caption{Comparison of photometric-metallicity estimates with those from the high-resolution spectroscopic metallicities for dwarf (red squares) and giant (blue squares) stars. The HRS samples we compare with include a sample of the most metal-poor stars \citep{2019ApJ...879...37N}, the $R$-Process Alliance sample \citep[RPA;][]{2018ApJ...858...92H, 2018ApJ...868..110S, 2020ApJ...898..150E, 2020ApJS..249...30H} for over 600 VMP stars, the CFHT ESPaDOnS follow-up observations of 132 metal-poor candidates selected from the Pristine survey \citep{2022MNRAS.511.1004L}, the Subaru follow-up observations of 400 VMP candidates selected from the LAMOST \citep{2022ApJ...931..146A, 2022ApJ...931..147L}, and the GTC follow-up observations of extremely metal-poor (EMP) candidates identified from the Pristine and LAMOST surveys \citep{2023MNRAS.519.5554A}.
The small and large circles mark stars with carbon enhancements ($+0.6 < $ [C/Fe] $ < +2.0$) and extreme carbon enhancements ([C/Fe]\,$ > +2.0$), respectively.
The central dashed line represents the one-to-one line; the other two dashed lines have shifts of $\pm 0.5$\,dex.}
\end{center}
\end{figure}

\begin{figure*}
\begin{center}
\includegraphics[scale=0.55,angle=0]{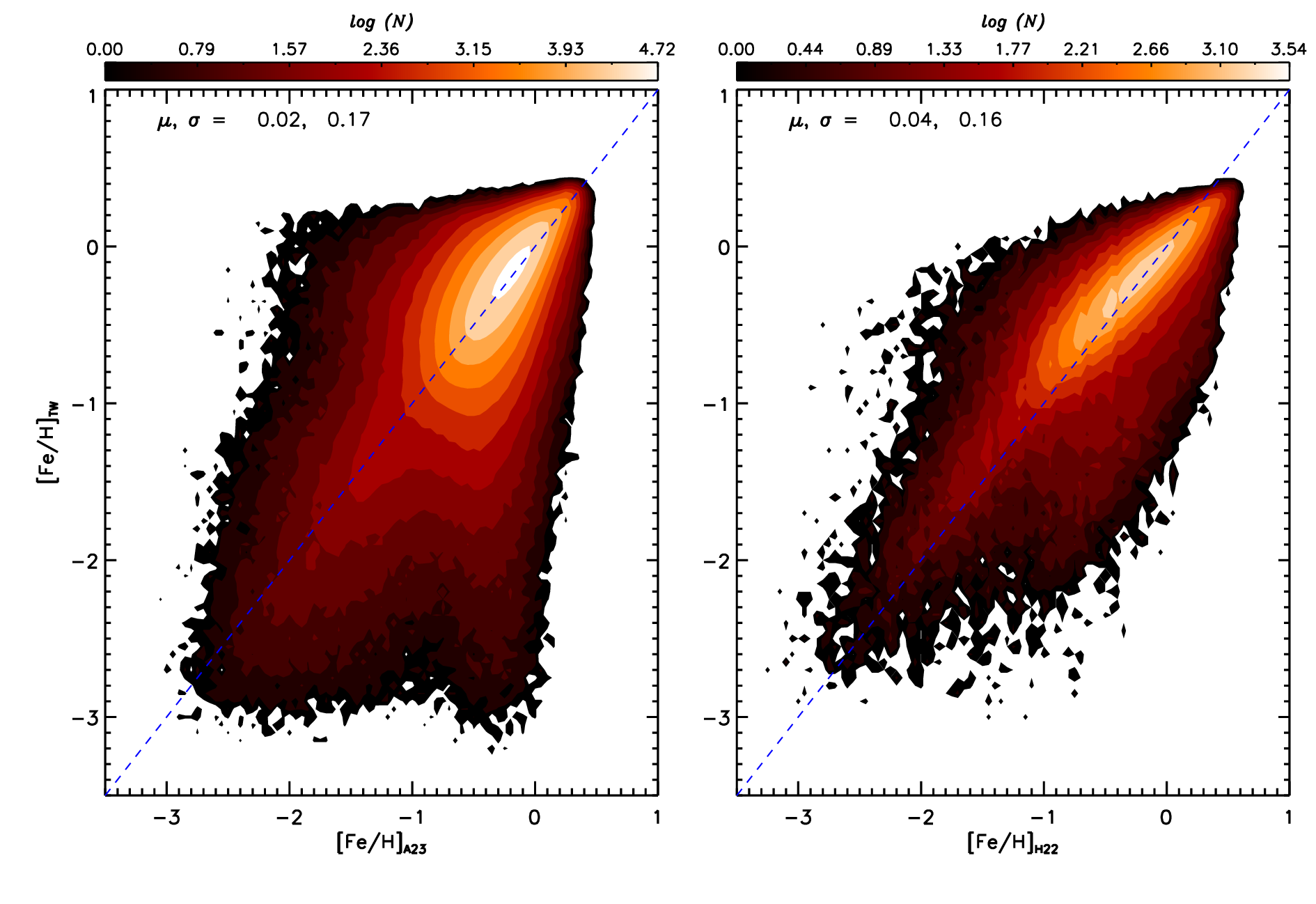}
\caption{Comparison of estimated photometric metallicity, [Fe/H], from this
work with that derived by \citet[][A23]{2023arXiv230202611A} for over 5 million stars in common (left panel), and that derived by H22 for about 390,000 stars in common (right panel). The color-coded contours of the stellar number density are shown with color bars on the top of each panel.
The values of median and standard deviation of the metallicity differences
(this work minus A22/H22) are marked in the top-left corners.}
\end{center}
\end{figure*}

\begin{figure*}
\begin{center}
\includegraphics[scale=0.425,angle=0]{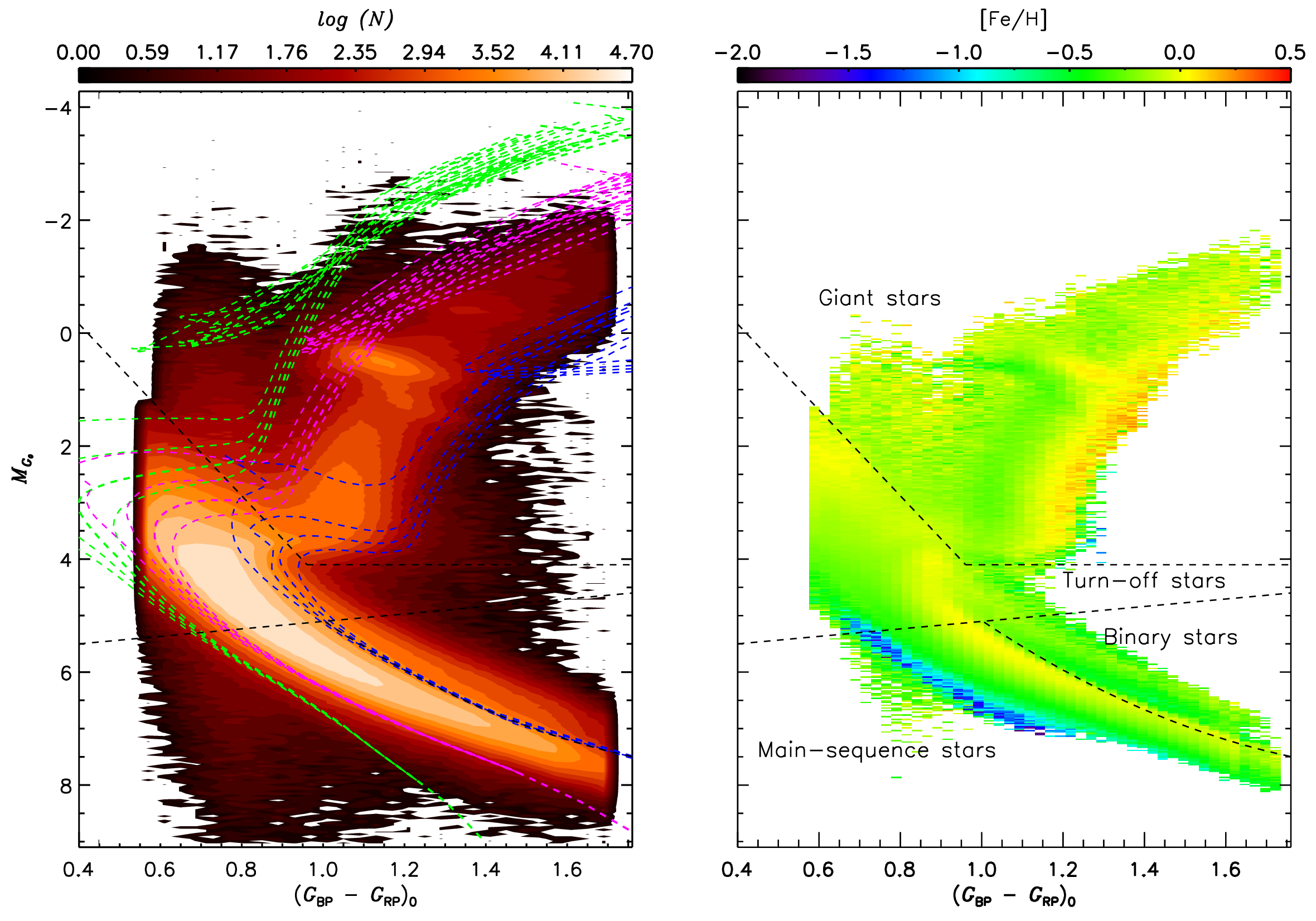}
\caption{{\it Left panel:} Number-density distribution (on a logarithmic scale) in the H-R diagram: $M_{G_0}$ versus $(G_{\rm BP} - G_{\rm RP})_0$.
The blue, magenta, and green dashed lines represent stellar isochrones from PARSEC \citep{2012MNRAS.427..127B, 2017ApJ...835...77M} with total metallicity [M/H]$= +0.50$, $-0.75$, and $-2.00$.
Isochrones of the same colors have different ages, from 3 to 9 Gyr, in steps of 2 Gyr (from left to right). 
The upper black dashed lines (defined in Fig.\,1) are used to classify dwarf and giant stars. 
The middle dashed line marks the main-sequence turn-off stars (above this line). 
The lower dashed line separates main-sequence (left part) and likely binary stars (right part).
{\it Right panel:} Similar to the left panel, but color-coded with the median photometric [Fe/H], as indicated by the top color bar.}
\end{center}
\end{figure*}

\begin{figure*}
\begin{center}
\includegraphics[scale=0.355,angle=0]{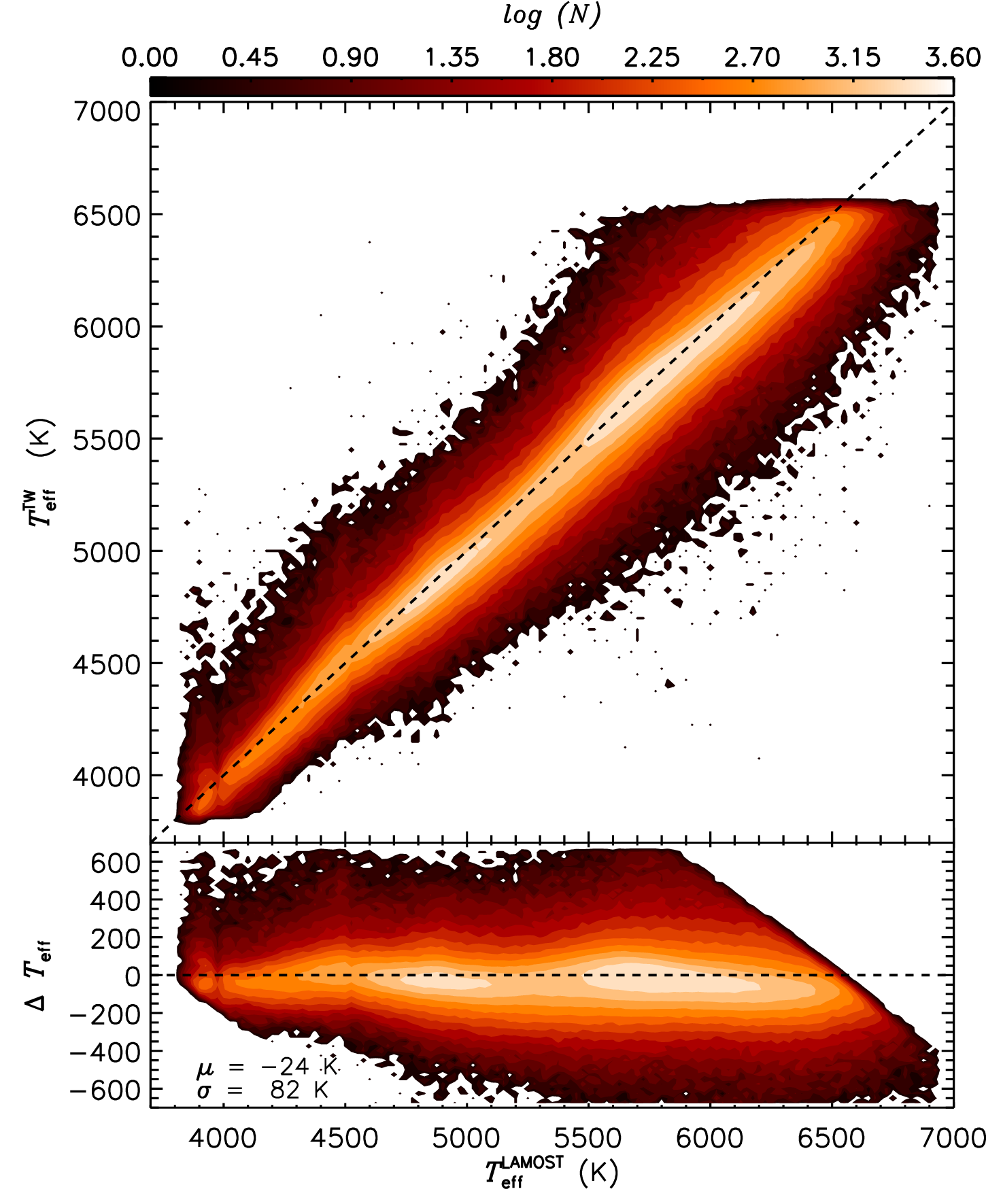}
\includegraphics[scale=0.355,angle=0]{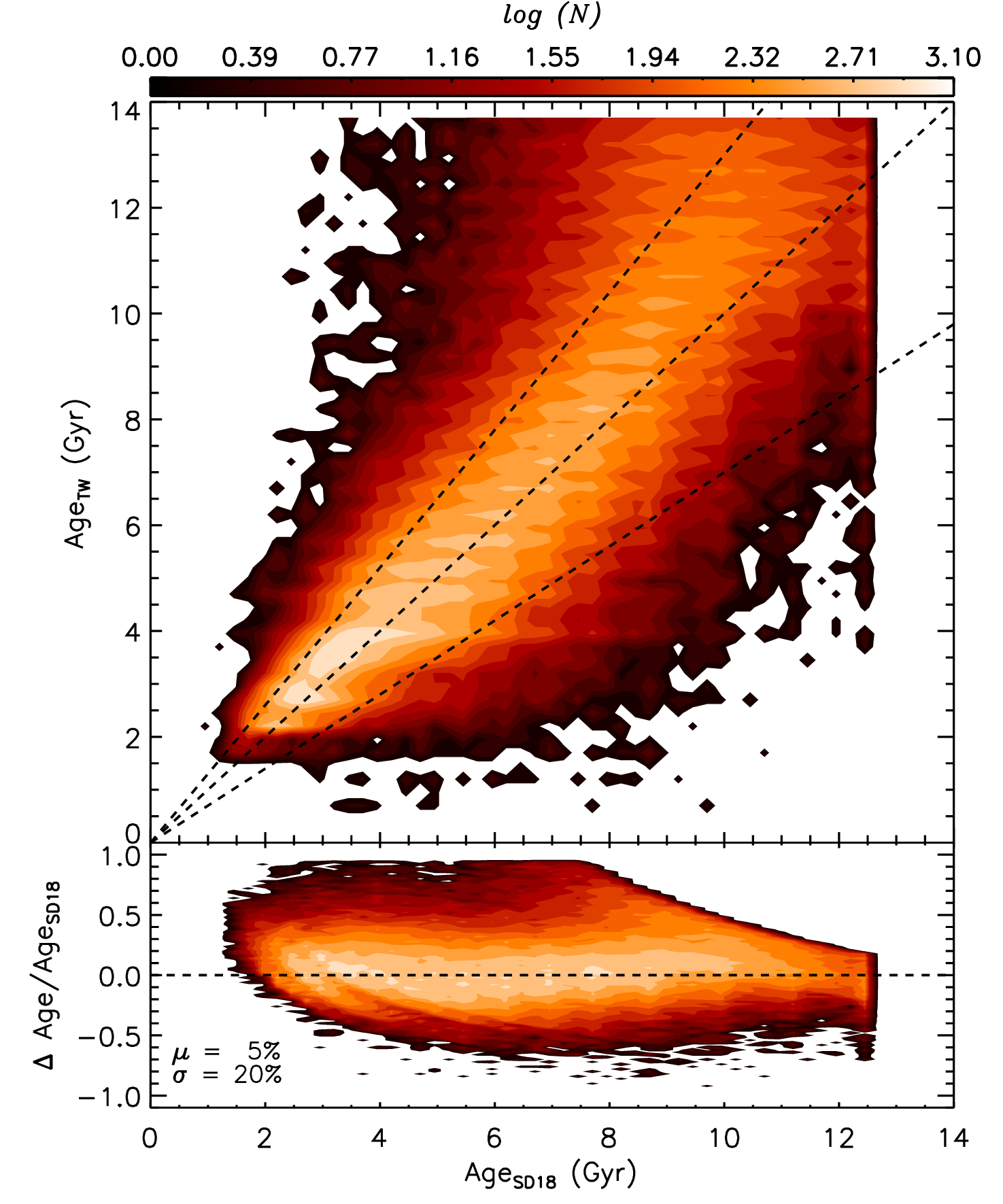}
\caption{ {\it Left panel:} Comparisons of effective temperature measurements between this work and LAMOST for over 159,000 stars in common.
The lower panel shows the effective temperature
difference (this work minus LAMOST), as a function of LAMOST effective temperature, with the values of the median and standard deviation of the effective temperature difference marked in the bottom left corner.
{\it Right panel:} Comparisons of stellar-age estimates between this work and \citet[][SD18]{2018MNRAS.481.4093S} for over 160,000 main-sequence turn-off stars in common. The dashed lines indicate Age$_{\rm TW} = 0.7, 1.0$ and $1.3$ times of Age$_{\rm SD18}$.
The lower panel shows the relative age
difference (this work minus SD18), as a function of SD18 age, with the values of the median and standard deviation of the relative age difference marked in the bottom left corner. In each panel, a color-coded contour of the stellar number density on a logarithmic scale is shown.}
\end{center}
\end{figure*}

\begin{figure}
\begin{center}
\includegraphics[scale=0.355,angle=0]{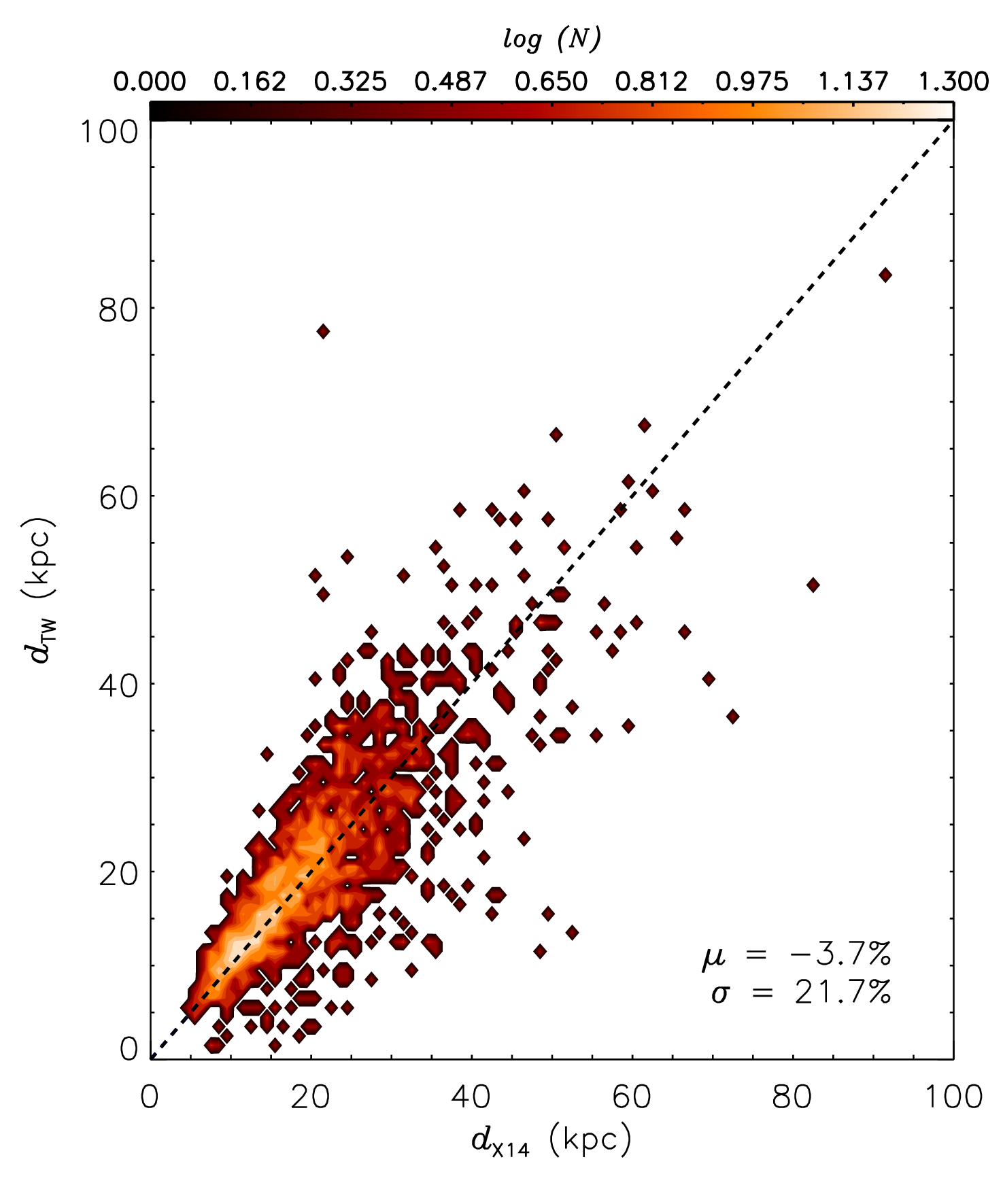}
\caption{Comparison of photometric distances estimated using the likelihood method in this work to those from \citet{2014ApJ...784..170X}.
The color-coded contour of the stellar number density is shown with color bars on the top of each panel.
The overall relative median offset and standard deviation of the distance difference are marked in the bottom-right corner.}
\end{center}
\end{figure}

\section{Data}
In the present work, the SAGES DR1 \citep{2023arXiv230615611F} dataset is adopted.
SAGES DR1 has released a total of about 100 million sources extracted from 36,092 accepted frames in the $uv$-bands collected by the 90-inch (2.3m) Bok Telescope at Kitt Peak National Observatory in Arizona.
DR1 covers about half of the Northern Hemisphere (9960 square degrees), about 90 per cent of the planned area.
The median completeness is about 20.4 and 20.3 for the $u$- and $v$-band, respectively.
This is one of the deepest near-ultraviolet large-scale photometric survey with a $5\sigma$ depth close to 21.5 in the $u$-band. 
Compared to other near-ultraviolet deep photometric surveys, e.g., the SDSS \citep{2000AJ....120.1579Y} and the South Galactic Cap $u$-band Sky Survey \citep[SCUSS;][]{2016RAA....16...69Z}, SAGES has the advantage of using the two medium-bandwidth filters $uv$, which are optimized for estimates of stellar parameters.

In addition to the $uv$-band data provided by SAGES DR1, the optical bands of $G, G_{\rm BP}, G_{\rm RP}$, as well as astrometric information, is adopted from the {\it Gaia} EDR3 \citep{2021A&A...649A...1G}. 
The {Gaia} EDR3 broadband photometry is essentially complete between $G = 12$ and $G = 17$. 
The completeness is quite complicated for sources fainter than $G = 17$, which is strongly dependent on celestial position \citep{2021A&A...649A...3R, 2023A&A...669A..55C, 2023arXiv230317738C}.
In total, nearly 33 million stars are selected by the following cuts:
 \begin{enumerate}[label=\arabic*)]

\item ${\rm flag}\_u/v = 0$ in SAGES DR1

\item Uncertainties of $G$, $G_{\rm BP}$, and $G_{\rm RP}$ smaller than 0.05\,mag

\item Galactic latitude $|b| \ge 10^{\circ}$
\end{enumerate}
SAGES was initially designed to avoid the high-reddening regions with $|b| \le 10^{\circ}$, although a few disk areas are observed for specific reasons.
The former two cuts are required for precise metallicity estimates, but they do affect the completeness in the faint range ($G > 18.5$).
The last cut is to exclude those disk regions in our analysis, given their high values of extinction.
This sample is referred to as the main sample for our following analysis.

In this study, the colors $u-G_{\rm BP}$, $v-G_{\rm RP}$, and $G_{\rm BP} - G_{\rm RP}$ are used.
We note that the mean $G_{\rm BP}$ flux in {\it Gaia} EDR3 is 
over-estimated for faint red sources with $G \ge 20.0$ \citep[e.g.,][]{2021A&A...649A...3R, 2022MNRAS.511..572O}. However, only 650 thousand stars (no more than 3 per cent of the full sample) in our final catalog are fainter than 20th magnitude in the $G$-band.
Therefore, the systematic issue for $G_{\rm BP}$ is minor for the current study.
Unless indicated otherwise, these colors are corrected for reddening using the extinction map of \citet[][hereafter SFD98]{1998ApJ...500..525S}
\footnote{Here the SFD98 $E(B-V)$ is corrected for a 14\% systematic 
over-estimate \citep[e.g.,][]{2011ApJ...737..103S, 2013MNRAS.430.2188Y}}.
The reddening coefficients for those colors, as well as for the $G$-band, are calculated using the same way as in H22.

\section{Metallicity Determination}

\subsection{Training Set}
The key to determinations of metallicity using stellar colors is the training set.
The training set adopted here is similar to that used in H22, which consists of 1) LAMOST DR9\footnote{\url{http://www.lamost.org/dr9/v1.0/}}, 2) the revised parameters of metal-poor ([Fe/H]\,$\leq -1.8$) stars of SEGUE \citep{2009AJ....137.4377Y, 2022ApJS..259...60R}, along with other datasets from SDSS (we refer to the total dataset below as SEGUE), and LAMOST \citep{2012RAA....12..735D, 2014IAUS..298..310L} by a custom version of the SSPP (LSSPP; Lee et al. 2015), along with careful visual inspection (by Beers), and 3) the bibliographical compilation of measurements
of stellar atmospheric parameters from high-resolution spectroscopy (HRS) by PASTEL \citep{2016A&A...591A.118S} and SAGA \citep{2008PASJ...60.1159S} .
The metallicity scale of the former two sets is calibrated to the one obtained from the HRS dataset.
More details of our efforts to construct a training set with a homogenous scale of metallicity, as well as other elemental-abundance ratios, will be described in Huang et al. (2023).

We then cross-match the above training set to the main sample, together with the following cuts:
 \begin{enumerate}[label=\arabic*)]

\item The stars must have small values of extinction (to minimize uncertainties due to reddening corrections): Galactic latitude $|b| \ge 20^{\circ}$ and $E (B - V) \leq 0.08$ 

\item The stars must have reliable metallicity estimates: LAMOST/SEGUE spectral signal-to-noise ratio (SNR) greater than 20, effective temperatures in the range 3800 $\leq T_{\rm eff}$\,${\rm (K)} \leq$ 7500 (i.e., typical FGK-type stars)

\item The photometric uncertainties in the SAGES $uv$ and {\it Gaia} $G_{\rm BP}G_{\rm RP}G$ bands must be smaller than 0.035\,mag

\item The stars must have $Gaia$ relative parallax measurement uncertainties smaller than 50\%

\end{enumerate}
In addition to the above cuts, only about half of the metal-rich ([Fe/H]\,$>-1.0$) stars are selected to avoid large differences in the number of metal-rich ([Fe/H]\,$>-1.0$)  and metal-poor ([Fe/H]\,$<-1.0$)  stars (see the right panel of Fig.\,1).
Given the number of stars in common between SAGES and those with spectroscopy, the cut on Galactic latitude would not introduce bias in the training sets, e.g., a lack of metal-rich disk populations (see the right panel of Fig.\,1).
A total of 223,537 stars (182,035 dwarfs and 41,502 giants) are selected to construct the final training set.
The absolute $G$-band magnitudes of these stars are derived by adopting the distances from \citet{2021AJ....161..147B}, based on the parallax measurements from {\it Gaia} EDR3.
The Hertzsprung–Russell (H-R) diagram of the training set is then shown in the left panel of Fig.\,1.
By using empirical cuts defined in H22, the training stars are further divided into dwarf and giant stars.
The right panel of Fig.\,1 shows the metallicity distributions of the dwarf and giant stars in the training set.

\subsection{Metallicity Estimation}
To estimate photometric metallicity, we first define the metallicity-dependent stellar loci of $(u/v-G_{\rm BP})_0$ versus $(G_{\rm BP} - G_{\rm RP})_0$ in Fig.\,2 for  both dwarf stars (top panel) and giant stars (bottom panel).
Similar to our results with SMSS DR2 in H22, both  $(u-G_{\rm BP})_0$ and $(v-G_{\rm BP})_0$ colors exhibit significant sensitivities to stellar metallicity for different types of stars characterized by $(G_{\rm BP} - G_{\rm RP})_0$.
Third-order 2D polynomials with 10 free parameters are then applied to describe the stellar loci of dwarf and giant stars:
 \begin{equation}
 \begin{split}
 (u/v - G_{\rm BP})_0 & =  a_{0,0} + a_{0,1}y + a_{0,2}y^2 + a_{0,3}y^3 + a_{1,0}x + \\
 & a_{1,1}xy + a_{1,2}xy^2 + a_{2,0}x^2 + a_{2,1}x^2y + a_{3,0}x^3\text{,}
 \end{split}
 \end{equation}
where $x$ and $y$ represent $(G_{\rm BP} - G_{\rm RP})_0$ and [Fe/H], respectively.
Two to three sigma-clipping is applied in the fitting process.
The resultant fit coefficients are listed in Table\,1.

 \begin{table}
\centering
\begin{threeparttable}
\caption{Fit Coefficients for Metallicity  Estimates for Dwarf and Giant Stars}
\begin{tabular}{c|rr|rrr}
\hline
\multirow{2}{*}{Coeff.} & \multicolumn{2}{c}{Dwarf Stars}&\multicolumn{2}{|c}{Giant Stars}\\
&$(u - G_{\rm BP})_0$\tnote{a}&$(v - G_{\rm BP})_0$\tnote{b}&$(u - G_{\rm BP})_0$\tnote{a}&$(v - G_{\rm BP})_0$\tnote{b}\\
\hline
$a_{0,0}$&$  3.084320$&$  0.694670$&$  3.603653$&$  0.372304$&\\
$a_{0,1}$&$ -0.587864$&$ -0.067923$&$ -1.418031$&$ -0.185474$&\\
$a_{0,2}$&$  0.016439$&$  0.097577$&$  0.000912$&$  0.087328$&\\
$a_{0,3}$&$  0.022031$&$  0.008902$&$  0.026515$&$ -0.001555$&\\
$a_{1,0}$&$ -8.246335$&$ -2.465311$&$ -9.413692$&$ -1.103181$&\\
$a_{1,1}$&$  1.795641$&$  0.589683$&$  3.089203$&$  0.830962$&\\
$a_{1,2}$&$  0.160856$&$ -0.043831$&$  0.178593$&$ -0.059142$&\\
$a_{2,0}$&$ 10.296704$&$  4.577793$&$ 10.752706$&$  2.737885$&\\
$a_{2,2}$&$ -0.676539$&$ -0.262572$&$ -1.183446$&$ -0.370985$&\\
$a_{3,0}$&$ -3.201283$&$ -1.498611$&$ -2.972827$&$ -0.727921$&\\
\hline

\hline
\end{tabular}
\begin{tablenotes}
\item[a]$(u - G_{\rm BP})_0 = a_{0,0} + a_{0,1}y + a_{0,2}y^2 + a_{0,3}y^3 + a_{1,0}x + a_{1,1}xy + a_{1,2}xy^2 + a_{2,0}x^2 + a_{2,1}x^2y + a_{3,0}x^3$, where $x$ and $y$ represent $(G_{\rm BP} -  G_{\rm RP})_0$ and [Fe/H], respectively.
\item[b]$(v - G_{\rm BP})_0 =  a_{0,0} + a_{0,1}y + a_{0,2}y^2 + a_{0,3}y^3 + a_{1,0}x + a_{1,1}xy + a_{1,2}xy^2 + a_{2,0}x^2 + a_{2,1}x^2y + a_{3,0}x^3$, where $x$ and $y$ represent $(G_{\rm BP} -  G_{\rm RP})_0$ and [Fe/H], respectively.
\end{tablenotes}
\end{threeparttable}
\end{table}

Using the stellar loci, one can determine the photometric metallicity using the maximum-likelihood approach developed in H22.
For a given star, the metallicity is obtained from the probability distribution function (PDF) of [Fe/H] estimated from the likelihood function:
 \begin{equation}
 L_c = \frac{1}{\sqrt{2\pi}\sigma_{c_{\rm obs}}}\exp{\frac{-(c_{\rm obs} - c_{\rm pred})^2}{2\sigma_{c_{\rm obs}}^2}},
 \end{equation}
where $c_{\rm obs}$ are the observed colors, i.e., $(u/v - G_{\rm BP})_0$, with assumed Gaussian errors $\sigma_{c_{\rm obs}}$.
The $c_{\rm pred}$ represents the same colors predicted by the metallicity-dependent stellar loci (defined by Equation\,1) with $(G_{\rm BP} - G_{\rm RP})_0$ from observations and [Fe/H] ranging from $-3.5$ to $+0.8$ in steps of 0.01\,dex.
The uncertainty in the photometric metallicity estimated is taken to be half of the 68\% interval of the resultant PDF.

From the above approach, we estimate the photometric metallicities of training-set stars to be compared to the spectroscopic measurements as an internal test.
These comparisons are shown in Fig.\,3 for both dwarf stars (top panel) and giant stars (bottom panel).
Generally, the estimated photometric metallicities agree with the spectroscopic metallicities very well for both dwarf and giant stars, either from $(u - G_{\rm BP})_0$ or $(v - G_{\rm BP})_0$; the overall scatter is only 0.09\,dex  and 0.13\,dex for dwarf stars achieved by $(u - G_{\rm BP})_0$ and $(v - G_{\rm BP})_0$, respectively.
The scatter of the combined estimates using an error-weighted mean is further reduced to 0.08\,dex, even better than the precision of low/medium-resolution spectroscopy.
As shown in the top-right panel of Fig.\,4, no significant systematic offset is found for dwarf stars with photometric [Fe/H]\,$>-1.0$, and a mild offset of $-0.20$ to $-0.4$\,dex (photometric minus spectroscopic) is found for metal-poor dwarf stars with photometric [Fe/H]\,$\le-1.0$.
The metallicity precision for dwarf stars as revealed by the internal comparisons is a function of [Fe/H], with scatter smaller than 0.1\,dex for [Fe/H]\,$>-0.5$, increasing to 0.3-0.4\,dex at the extremely metal-poor end ([Fe/H]\,$\sim -3.0$).
For giant stars, the overall scatter is around 0.11\,dex.
The comparisons show that photometric metallicity derived from $(v - G_{\rm BP})_0$ is in excellent agreement with that of spectroscopy, with negligible offsets for [Fe/H]\,$>-2.0$ and a small offset of $-0.2$\,dex (photometric minus spectroscopic) at the extremely metal-poor end  ([Fe/H]\,$\sim -3.0$).
The metallicity precision from $(v - G_{\rm BP})_0$ is around 0.1\,dex for [Fe/H]\,$>-1.0$, and 0.2-0.3\,dex for [Fe/H]\,$\le -1.0$.
The performance of photometric metallicity derived from $(u - G_{\rm BP})_0$ is moderately worse, especially for warmer giant stars, which are mostly BHB stars (see the blue box in the bottom left panel of Fig.\,3).
Finally, the internal checks indicate that there are no systematic trends with effective temperature for the photometric-metallicity estimates of both dwarf and giant stars (see the top-left panel of Fig.\,4).

In addition to the internal test, we derive photometric metallicities for LAMOST targets with larger values of $E (B-V)$ that are not included in the training set.
Using the LAMOST targets (including these stars with low values of extinction in the training set), we show the metallicity differences between the photometric and spectroscopic values as a function of $E (B-V)$ in Fig.\,5.
The metallicity differences (photometric minus spectroscopic) steadily decrease with  $E (B-V)$, and reach $\sim +0.2$\,dex at $E (B-V) \sim 0.5$ for both dwarf and giant stars.
This trend is possibly due to the spatial systematic uncertainties of theSFD98 extinction map, as found most recently by  \citet{2022ApJS..260...17S}.
Moreover, \citet{2023ApJS..264...14Z} have shown that the reddening coefficients depend not only on effective temperature/intrinsic colors, but also extinction itself (ignored in this work).
The neglect of the extinction term may also partly contribute to this $E (B-V)$ dependent trend.
To correct for this systematic trend, a fifth-order polynomial is applied to describe the differences as a function of $E (B-V)$ for dwarf and giant stars, respectively.

According to the above tests, the final metallicity of a dwarf star is given by the combined estimate if both $(u - G_{\rm BP})_0$ and $(v - G_{\rm BP})_0$ colors are
available, or given by the single measurement from either $(u - G_{\rm BP})_0$ or $(v - G_{\rm BP})_0$, depending on which color is available. 
The final metallicity of a giant star is given by the measurement of color $(v - G_{\rm BP})_0$, or the color $(u - G_{\rm BP})_0$ if the former is not available.
In this manner, photometric-metallicity estimates are derived for over 26 million stars (23 million dwarf stars and 3 million giant stars) in SAGES.
Note that the extinction dependent zero-point offsets are corrected using the fifth-order polynomial constructed above. 
The $G$-band magnitude distributions of stars with metallicity estimates are shown in the left panel of Fig.\,6.

The overall completeness limit is around magnitudes $G$ = 17.5 and 18.5, for dwarf and giant stars, respectively.
As mentioned earlier, we caution that the completeness of {\it Gaia} broadband photometry is quite complicated, especially in crowded regions, for stars with $G > 17$ \citep{2021A&A...649A...3R, 2023A&A...669A..55C, 2023arXiv230317738C}.
The photometric-metallicity distributions of dwarf and giant stars are shown in the right panel of Fig.\,6.
The total number of very metal-poor (VMP; [Fe/H]\,$< -2.0$) stars is about one million, which is the largest database of VMP candidates yet assembled from photometric techniques.

The metallicity uncertainty of a star is contributed by two sources: the method error deduced from the internal checks and the random errors derived from the likelihood function.
The metallicity uncertainty as a function of $G$-band magnitude is shown in Fig.\,7, which is dominated by the method error and random errors in the bright and faint end, respectively.

\subsection{Comparison with APOGEE DR17 and GALAH DR3+}
The accuracy of our  photometric estimates of metallicity is examined by comparisons with the independent spectroscopic measurements from the APOGEE DR17 \citep{2022ApJS..259...35A} and GALAH DR3+ \citep{2021MNRAS.506..150B}.
The comparisons are shown in Fig.\,8 for 72,995 high-quality (SNR\,$\ge 30$) stars in common with APOGEE  and 13,038 high-quality (SNR\,$\ge 30$) stars in common with GALAH DR3+.
Generally, the photometric-metallicity estimates agree very well with the spectroscopic values, without significant offsets.
The overall scatter is only 0.09\,dex for dwarf stars and 0.10-0.15\,dex for giant stars. 
The zero-point and precision of individual metallicity bins are also examined in the lower panels of Fig.\,8; the results are consistent with our internal tests (see Fig.\,4).

We also present the metallicity differences between the photometric estimates and spectroscopic values from APOGEE DR17 as a function of $E (B-V)$ in Fig.\,9.
The plot clearly shows that the offsets are all around zero for different bins of $E (B-V)$, a validation of our polynomial corrections described in Section\,3.2 (see Fig.\,5).

 \begin{table*}
\centering
\caption{Sample Content}
\begin{tabular}{ccccccc}
\hline
\hline
&Dwarf &Giant&All\\
\hline
\textbf{Total}&\textbf{22,529,640}&\textbf{3,202,065}&\textbf{25,731,705}\\
\hline
\textbf{Stars with [Fe/H] measurements}&\textbf{22,529,640}&\textbf{3,202,065}&\textbf{25,731,705}\\
\text{[Fe/H]} measured by $u - G_{\rm BP}$&5,955,809&814,066&6,769,875\\
\text{[Fe/H]} measured by $v - G_{\rm BP}$&2,611,929&2,387,999&4,999,928\\
\text{[Fe/H]} measured by two colors&13,961,902&--&13,961,902\\
\hline
\textbf{Stars with $T_{\rm eff}$ measurements}&\textbf{22,529,640}&\textbf{3,202,065}&\textbf{25,731,705}\\
\hline
\textbf{Stars with distance measurements}&\textbf{14,690,031}&\textbf{2,951,337}&\textbf{17,641,368}\\
Distance estimated by Gaia EDR3 parallax&13,595,147&1,409,144&15,004,291\\
Distance estimated by color-absolute magnitude fiducials&1,094,884&1,542,193&2,637,077\\
\hline
\textbf{Stars with age measurements}&\textbf{13,428,487}&\textbf{1,365,693}&\textbf{14,794,180}\\
\hline
\textbf{Stars with RV measurements}&\textbf{3,045,883}&\textbf{1,168,081}&\textbf{4,213,964}\\
RV measured from GALAH DR3+&24,261&11,918&36,179\\
RV measured from APOGEE DR17&43,983&56,927&1,100,910\\
RV measured from Gaia DR3&2,038,911&1,001,874&3,040,785\\
RV measured from RAVE DR5&25&11&36\\
RV measured from LAMOST DR9&836,516&87,902&924,418\\
RV measured from SDSS/SEGUE DR16&102,187&9449&111,636\\
\hline
\end{tabular}
\end{table*}

\subsection{Comparison with Metal-poor Samples from High-resolution Spectroscopy}
To explore the capabilities of the SAGES filters for determinations of metallicity for metal-poor stars, we collect samples of independent metallicity estimates from HRS, especially for metal-poor stars.
The HRS samples we compare with include a sample of the most metal-poor stars \citep{2019ApJ...879...37N}, the $R$-Process Alliance sample \citep[RPA;][]{2018ApJ...858...92H, 2018ApJ...868..110S, 2020ApJ...898..150E, 2020ApJS..249...30H} for over 600 VMP stars, the CFHT ESPaDOnS follow-up observations of 132 metal-poor candidates selected from the Pristine survey \citep{2022MNRAS.511.1004L}, the Subaru follow-up observations of 400 VMP candidates selected from the LAMOST \citep{2022ApJ...931..146A, 2022ApJ...931..147L}, and the GTC follow-up observations of extremely metal-poor (EMP) candidates identified from the Pristine and LAMOST surveys \citep{2023MNRAS.519.5554A}.

We cross-match the SAGES sample to the collected HRS samples and find 112 stars in common (54 dwarfs and 58 giant stars).
The comparison result is shown in Fig.\,10.
Generally, our photometric-metallicity estimates are consistent with the HRS values for metal-poor stars without significant carbon enhancements ([C/Fe]\,$< +0.6$).
The overall scatter of the differences (photometric minus spectroscopic) is 0.57\,dex and 0.30\,dex, respectively, for dwarf and giant stars, with mild offsets of $+0.38$\,dex and $+0.18$\,dex, respectively .
The result is in line with our internal checks (see Fig.\,4).
We note the photometric-metallicity estimates of ultra metal-poor (UMP; [Fe/H] $< -4.0$) stars can be over-estimated by up to 2\,dex for stars with very high carbon enhancements ([C/Fe]\,$\ge +2.0$).

\subsection{Comparison with SMSS and Gaia XP Spectra}
We compare our results to those of H22 from SMSS and those of \citet{2023arXiv230202611A} from Gaia XP low-resolution spectra.  The latter has recently delivered estimates of metallicity using a 
data-driven technique for over 120 million stars from Gaia XP low-resolution spectra.
As shown in Fig.\,11, our estimates are consistent with those of \citet{2023arXiv230202611A} and H22, with tiny offsets and a scatter smaller than 0.20\,dex.
Finally, although the total number of our metallicity estimates (SAGES + SMSS) does not exceed 50 million stars,
we emphasize that the volume of our sample is much larger than that of sample constructed from Gaia XP spectra, given that the limiting magnitude of SAGES and SMSS is nearly 3\,mag deeper than that of the Gaia XP spectra.
This larger volume will enable numerous interesting studies of the Milky Way, e.g., searching for substructures in the stellar halo.

\section{Effective temperature, Distance, and Age Estimates}
The effective temperatures of dwarf and giant stars are derived from the metallicity-dependent $T_{\rm eff}$--color relations constructed in H22.
Here the color is the de-reddened $(G_{\rm BP} - G_{\rm RP})_0$, and metallicity is given by photometric [Fe/H].
In this way, effective temperatures are obtained for all of our program stars.
As examined with over 159,000 stars in common, the effective temperature estimated in this work is quite consistent with that from LAMOST, with a small offset around $-24$\,K (this work minus LAMOST) and a scatter of only 84\,K (see Fig.\,13).

Distances estimated by \citet{2021AJ....161..147B} are adopted for stars with reliable parallax measurements with precision better than 30\%, parallax greater than 0.15\,mas, and  renormalized unit weight error (RUWE) smaller than 1.4.
A total of 15,974,812 stars have distances estimated in this way.
Using the apparent $G$-band magnitudes and SFD $E (B-V)$, the $G$-band absolute magnitudes have been derived for the nearly 16 million stars with reliable geometric distances.
Fig.\,12 is the Hertzsprung-Russell (H-R) diagram for about 8 million stars with relative parallax error better than 10\%, parallax greater than 0.4\,mas, and RUWE$\le 1.4$.
Guided by the isochrones  of PARSEC  \citep{2012MNRAS.427..127B, 2017ApJ...835...77M}, empirical cuts are defined to further classify dwarf stars into main-sequence turn-off, main-sequence,
and binary stars. 

For the stars without geometric distance estimates, the distances are obtained by inferring their absolute magnitudes from the constraints of stellar colors and photometric metallicity.
For main-sequence dwarf stars,  the $G$-band absolute magnitudes are derived from the third-order 2D polynomial relation constructed in H22.
Combining with the $G$-band magnitude and the SFD $E (B -V)$, the distances are found for over one million main-sequence dwarf stars with $(G_{\rm BP} - G_{\rm RP})_0 \ge 1.0$.
For giant stars, a likelihood method developed in \citet[][hereafter X14]{2014ApJ...784..170X} and \citet{2019ApJS..243....7H} is adopted to infer the $i$-band absolute magnitude using the $(g - i)_0$ color, photometric [Fe/H], and empirical color--magnitude fiducials interpolated from six globular clusters.
Here, the $g$- and $i$-band magnitudes are from the Pan-STARRS1 surveys \citep[PS1;][]{2016arXiv161205560C}; the reddening-correction coefficients are from \citet{2019ApJ...887...93G}.
The interested reader is referred to X14 or \citet{2019ApJS..243....7H} for more details. 

In the above manner, a total of over 1.6 million giant stars have their distances estimated. To test the accuracies of our distance estimates for giant stars, Fig.\,14 compares these with those of X14 for over 1600 stars in common.
The results are consistent with each other, with a tiny relative offset of $-3.7$\% (this work minus X14) and a scatter of 21.7\%.
This scatter implies that both estimates have a typical precision of about 16\%, which is expected by X14.

Finally, we derive stellar ages for stars with good parallax measurements, i.e., parallax measurements with precision better than 30\%, parallax greater than 0.15\,mas, and RUWE$\le 1.4$, using the technique developed in H22.
Nearly 15 million stars have their ages estimated in this way.
We note that the RUWE cut cannot exclude all of the binary stars, whose ages may be over-estimated.
As noted by H22, this technique is mostly valid for main-dequence turn-off and sub-giant stars; uncertainties are larger for other types of stars in the H-R diagram.
We perform a similar check as done in H22 with over 160,000 stars in common between this work and \citet[][SD18]{2018MNRAS.481.4093S}, who derived isochrone ages for over 3 million stars with both spectroscopic and astrometric information.
The check shows that the age estimates in this work agree with with those from SD18, with an offset of 5\% in relative age difference (age$_{\rm TW} - $age$_{\rm SD18})$/age$_{\rm SD18}$ and a scatter in the relative age difference of around 20\%.

\begin{table*}
\centering
\caption{Description of the Final Sample}
\begin{tabular}{lll}
\hline
\hline
Field&Description&Unit\\
\hline
Sourceid&Gaia EDR3 source ID&--\\
ra&Right Ascension from SAGES DR1 (J2000)&degrees\\
dec&Declination from  SAGES DR1 (J2000)&degrees\\
gl&Galactic longitude derived from ICRS coordinates&degrees\\
gb&Galactic latitude derived from ICRS coordinates&degrees\\
u/v&Magnitudes for the SAGES two bands from SAGES DR1&--\\
err\_u/v&Uncertainties magnitudes for the SAGES two bands from SAGES DR1&mag\\
g/r/i&Magnitudes from Pan-STARRS1&--\\
err\_g/r/i&Uncertainties of magnitudes from Pan-STARRS1&mag\\
G/BP/RP&Magnitudes for the {\it Gaia} three bands from EDR3; note G represents a calibration-corrected G magnitude&--\\
err\_G/BP/RP&Uncertainties of magnitudes for the three {\it Gaia} bands from EDR3&mag\\
ebv\_sfd&Value of $E (B - V)$ from the extinction map of SFD98, corrected for a 14\% systematic&--\\
BR$_0$/uB$_0$/vB$_0$&Intrinsic colors of $(G_{\rm BP} - G_{\rm RP})_0$, $(u - G_{\rm BP})_0$, and $(v - G_{\rm BP})_0$&-- \\
err\_BR$_0$/uB$_0$/vB$_0$&Uncertainties of intrinsic colors of $(G_{\rm BP} - G_{\rm RP})_0$, $(u - G_{\rm BP})_0$, and $(v - G_{\rm BP})_0$&mag\\
\text{[Fe/H]}&Photometric metallicity&--\\
\text{err\_[Fe/H]}&Uncertainty of photometric metallicity&dex\\
\text{flg\_[Fe/H]}&Flag to indicate the stellar color(s) used in estimating [Fe/H], which takes the values ``ub", ``vb", and ``ub+vb"&--\\
$T_{\rm eff}$&Effective temperature&K\\
err\_$T_{\rm eff}$&Uncertainty of effective temperature&K\\
\text{dist\_adop}&Distance&kpc\\
\text{err\_dist\_adop}&Uncertainty of distance&kpc\\
\text{dist\_adop\_flg}&Flag to indicate the method used to derive distance, which takes the values ``parallax", ``CMF", and ``NO"&--\\
X/Y/Z&3D positions in the right-handed Cartesian system&kpc\\
err\_X/Y/Z&Uncertainties of 3D positions in the right-handed Cartesian system&kpc\\
$R_{\rm GC}$&Galactocentric distance&kpc\\
err\_$R_{\rm GC}$&Uncertainty of Galactocentric distance&kpc\\
$R$&Projected Galactocentric distance onto the Galactic plane&kpc\\
err\_$R$&Uncertainty of projected Galactocentric distance&kpc\\
\text{age}&Stellar age&Gyr\\
\text{err\_age}&Uncertainty of stellar age&Gyr\\
rv\_adop&Radial velocity&km s$^{-1}$\\
err\_rv\_adop&Uncertainty of radial velocity&km s$^{-1}$\\
rv\_adop\_flg&Flag to indicate the source of radial velocity, which takes the values ``GALAH", ``APOGEE'', ``Gaia",&--\\
& ``RAVE", ``LAMOST", ``SEGUE"&--\\
parallax&Parallax from {\it Gaia} EDR3&mas\\
err\_parallax&Uncertainty of parallax from {\it Gaia} EDR3&mas\\
pmra&Proper motion in Right Ascension direction from  {\it Gaia} EDR3&mas yr$^{-1}$\\
err\_pmra&Uncertainty of proper motion in Right Ascension direction from  {\it Gaia} EDR3&mas yr$^{-1}$\\
pmdec&Proper motion in Declination direction  from  {\it Gaia} EDR3&mas yr$^{-1}$\\
err\_pmdec&Uncertainty of proper motion in Declination direction from  {\it Gaia} EDR3&mas yr$^{-1}$\\
ruwe&Renormalised unit weight error from  {\it Gaia} EDR3&--\\
type&Flag to indicate classifications of stars, which takes the values ``dwarf" and ``giant" &--\\
subtype&Flag to indicate further sub-classifications of dwarf stars, which takes the values  ``TO", ``MS" and ``Binary"&--\\
\hline
\end{tabular}
\end{table*}

\section{Radial velocities and the final sample}
We collect measurements of radial velocities for our sample stars available from from completed and ongoing spectroscopic surveys, including 
GALAH DR3+ \citep{2021MNRAS.506..150B}, SDSS/APOGEE DR17 \citep{2022ApJS..259...35A}, {\it Gaia} DR3 \citep{2022arXiv220605902K}, RAVE DR5 \citep{2017AJ....153...75K}, LAMOST DR9\footnote{\url{http://www.lamost.org/dr9/v1.0/}} and SDSS/SEGUE DR16 \citep{2020ApJS..249....3A}, with typical measurement errors of 1.1, 0.5, 1.0-6.0, 2.0, 5.0 and 5.0\,km\,s$^{-1}$, respectively.
In total, over 4.2 million stars in our final sample have radial velocity measurements.
The detailed contributions of radial velocities from each survey are given in Table\,2.
If a star has radial velocity measurements from two more surveys, the result from the survey with the highest spectral resolution is adopted.
We note that all of the radial velocity zero-points are calibrated to the updated APOGEE radial-velocity standard stars based on the SDSS/APOGEE DR17 constructed using the same technique proposed in \citet{2018AJ....156...90H}.

In the final sample, over 22 million dwarf and 3 million giant stars have photometric-metallicity estimates (see Section\,3) from the stellar colors provided by SAGES DR1 \citep{2023arXiv230615611F} and {\it Gaia} EDR3 \citep{2021A&A...649A...1G}, and effective temperature estimates from the intrinsic $(G_{\rm BP} - G_{\rm RP})_0$ colors and photometric [Fe/H] (see Section\,4).
From the well-developed techniques described in H22, distances and ages are further derived for 18 and 15 million stars in the final sample, respectively (see Section\,4).
The radial velocity measurements, if available from the spectroscopic surveys, and the astrometric parameters in {\it Gaia} EDR3 \citep{2021A&A...649A...1G} are also included.

A description of the information for stars in the final sample catalog is presented in Table\,3.
The final stellar-parameter sample catalog will be released by the SAGES project as a value added catalog.
This sample already represents large progress on the development of stellar samples from the Northern sky for use in Galactic studies.
Together with our former effort from SMSS DR2 described in the first paper in this series, the sum of which represent photometric metallicities for on the order of 50 million stars, these results will shed light on understanding the formation and evolutionary history of our Galaxy.

The next step of this project is to extend this technique to derive photometric-metallicity with improved precision, especially at the metal-poor end, and other 
elemental-abundance ratios (e.g., [$\alpha$/Fe] and [C/Fe]) from the narrow/medium-band photometric surveys \citep[e.g., J/S-PLUS,][]{2019A&A...622A.176C, 2019MNRAS.489..241M}, or from {\it Gaia} XP low-resolution spectra, although only for stars with a relatively bright limiting magnitude around $G \sim 17.5$\,mag \citep{2022arXiv220606215G, 2023arXiv230202611A}.

\section{Summary}
In this, the second paper of this series, we present stellar parameters for over 20 million stars in the Northern sky, using SAGES DR1 and {\it Gaia} EDR3.
With a careful and comprehensive selection of a training set from spectroscopic measurements, we present photometric-metallicity estimates for nearly 26 million stars (23 million dwarf and 3 million giant stars), with useful metallicity determinations down to  [Fe/H] = $-3.5$.
Both internal and external checks show that the precisions of our photometric measurements are about $0.1$\,dex in the metal-rich range ([Fe/H]\,$> -1.0$) and  0.15-0.25/0.3-0.4\,dex for dwarf/giant stars with [Fe/H]$\le -1.0$.
This result is comparable to or even better than obtained for the low/medium-resolution spectroscopy.
In addition to metallicity, the final sample also includes measurements of effective temperature from metallicity-dependent $T_{\rm eff}$--color relations, distances  either from {\it Gaia} parallax measurements or from the metallicity-dependent color-absolute magnitude fiducials, and ages from comparisons between observations and stellar isochrones.
Radial velocities from spectroscopic surveys and astrometric parameters from {\it Gaia} EDR3 are also included.

To date, we have delivered stellar parameters for over 50 million stars covering almost 3$\pi$ steradians of sky, which will be useful to a variety of studies of the Milky Way.

 \section*{Acknowledgements} 
This work is supported by National Key R\&D Program of China No. 2019YFA0405500 and National Natural Science Foundation of China grants 11903027, 11833006, 11973001, 11603002, 11811530289 and U1731108. 
We used data from the European Space Agency mission Gaia (\url{http://www.cosmos.esa.int/gaia}), processed by the Gaia Data Processing and Analysis Consortium (DPAC; see \url{http://www.cosmos.esa.int/web/gaia/dpac/consortium}). 
T.C.B. acknowledges partial support from grant PHY 14-30152, Physics
Frontier Center/JINA Center for the Evolution of the
Elements (JINA-CEE), awarded by the US National Science
Foundation. His participation in this work was initiated by conversations that took place during a visit to China in 2019, supported by a PIFI Distinguished Scientist award from the Chinese Academy of Science.  Y.S.L. acknowledges support from the National Research Foundation (NRF) of
Korea grant funded by the Ministry of Science and ICT (NRF-2021R1A2C1008679).
Y.S.L. also gratefully acknowledges partial support for his visit to the University
of Notre Dame from OISE-1927130: The International Research Network for Nuclear Astrophysics (IReNA),
awarded by the US National Science Foundation.
CAO acknowledges support from the Australian Research Council through Discovery Project DP190100252.

The Stellar Abundance and Galactic Evolution Survey (SAGES) is a multi-band photometric project built and managed by the Research Group of the Stellar Abundance and Galactic Evolution of the National Astronomical Observatories, Chinese Academy of Sciences (NAOC). 

The national facility capability for SkyMapper has been funded through ARC LIEF grant LE130100104 from the Australian Research Council, awarded to the University of Sydney, the Australian National University, Swinburne University of Technology, the University of Queensland, the University of Western Australia, the University of Melbourne, Curtin University of Technology, Monash University and the Australian Astronomical Observatory. SkyMapper is owned and operated by The Australian National University's Research School of Astronomy and Astrophysics. The survey data were processed and provided by the SkyMapper Team at ANU. The SkyMapper node of the All-Sky Virtual Observatory (ASVO) is hosted at the National Computational Infrastructure (NCI). Development and support the SkyMapper node of the ASVO has been funded in part by Astronomy Australia Limited (AAL) and the Australian Government through the Commonwealth's Education Investment Fund (EIF) and National Collaborative Research Infrastructure Strategy (NCRIS), particularly the National eResearch Collaboration Tools and Resources (NeCTAR) and the Australian National Data Service Projects (ANDS).

The Guoshoujing Telescope (the Large Sky Area Multi-Object Fiber Spectroscopic Telescope, LAMOST) is a National
Major Scientific Project built by the Chinese Academy of Sciences. Funding for the project has been provided by the
National Development and Reform Commission. LAMOST is operated and managed by the National Astronomical Observatories, Chinese Academy of Sciences.

\bibliographystyle{apj}
\bibliography{BspII}

\begin{thebibliography}{}
\expandafter\ifx\csname natexlab\endcsname\relax\def\natexlab#1{#1}\fi

\bibitem[{{Abdurro'uf} {et~al.}(2022){Abdurro'uf}, {Accetta}, {Aerts}, {Silva
  Aguirre}, {Ahumada}, {Ajgaonkar}, {Filiz Ak}, {Alam}, {Allende Prieto},
  {Almeida}, {Anders}, {Anderson}, {Andrews}, {Anguiano}, {Aquino-Ort{\'\i}z},
  {Arag{\'o}n-Salamanca}, {Argudo-Fern{\'a}ndez}, {Ata}, {Aubert},
  {Avila-Reese}, {Badenes}, {Barb{\'a}}, {Barger}, {Barrera-Ballesteros},
  {Beaton}, {Beers}, {Belfiore}, {Bender}, {Bernardi}, {Bershady}, {Beutler},
  {Bidin}, {Bird}, {Bizyaev}, {Blanc}, {Blanton}, {Boardman}, {Bolton},
  {Boquien}, {Borissova}, {Bovy}, {Brandt}, {Brown}, {Brownstein}, {Brusa},
  {Buchner}, {Bundy}, {Burchett}, {Bureau}, {Burgasser}, {Cabang}, {Campbell},
  {Cappellari}, {Carlberg}, {Wanderley}, {Carrera}, {Cash}, {Chen}, {Chen},
  {Cherinka}, {Chiappini}, {Choi}, {Chojnowski}, {Chung}, {Clerc}, {Cohen},
  {Comerford}, {Comparat}, {da Costa}, {Covey}, {Crane}, {Cruz-Gonzalez},
  {Culhane}, {Cunha}, {Dai}, {Damke}, {Darling}, {Davidson}, {Davies},
  {Dawson}, {De Lee}, {Diamond-Stanic}, {Cano-D{\'\i}az}, {S{\'a}nchez},
  {Donor}, {Duckworth}, {Dwelly}, {Eisenstein}, {Elsworth}, {Emsellem},
  {Eracleous}, {Escoffier}, {Fan}, {Farr}, {Feng}, {Fern{\'a}ndez-Trincado},
  {Feuillet}, {Filipp}, {Fillingham}, {Frinchaboy}, {Fromenteau}, {Galbany},
  {Garc{\'\i}a}, {Garc{\'\i}a-Hern{\'a}ndez}, {Ge}, {Geisler}, {Gelfand},
  {G{\'e}ron}, {Gibson}, {Goddy}, {Godoy-Rivera}, {Grabowski}, {Green},
  {Greener}, {Grier}, {Griffith}, {Guo}, {Guy}, {Hadjara}, {Harding},
  {Hasselquist}, {Hayes}, {Hearty}, {Hern{\'a}ndez}, {Hill}, {Hogg},
  {Holtzman}, {Horta}, {Hsieh}, {Hsu}, {Hsu}, {Huber}, {Huertas-Company},
  {Hutchinson}, {Hwang}, {Ibarra-Medel}, {Chitham}, {Ilha}, {Imig}, {Jaekle},
  {Jayasinghe}, {Ji}, {Johnson}, {Jones}, {J{\"o}nsson}, {Katkov}, {Khalatyan},
  {Kinemuchi}, {Kisku}, {Knapen}, {Kneib}, {Kollmeier}, {Kong}, {Kounkel},
  {Kreckel}, {Krishnarao}, {Lacerna}, {Lane}, {Langgin}, {Lavender}, {Law},
  {Lazarz}, {Leung}, {Leung}, {Lewis}, {Li}, {Li}, {Lian}, {Liang}, {Lin},
  {Lin}, {Lin}, {Lintott}, {Long}, {Longa-Pe{\~n}a}, {L{\'o}pez-Cob{\'a}},
  {Lu}, {Lundgren}, {Luo}, {Mackereth}, {de la Macorra}, {Mahadevan},
  {Majewski}, {Manchado}, {Mandeville}, {Maraston}, {Margalef-Bentabol},
  {Masseron}, {Masters}, {Mathur}, {McDermid}, {Mckay}, {Merloni},
  {Merrifield}, {Meszaros}, {Miglio}, {Di Mille}, {Minniti}, {Minsley},
  {Monachesi}, {Moon}, {Mosser}, {Mulchaey}, {Muna}, {Mu{\~n}oz}, {Myers},
  {Myers}, {Nadathur}, {Nair}, {Nandra}, {Neumann}, {Newman}, {Nidever},
  {Nikakhtar}, {Nitschelm}, {O'Connell}, {Garma-Oehmichen}, {Luan Souza de
  Oliveira}, {Olney}, {Oravetz}, {Ortigoza-Urdaneta}, {Osorio}, {Otter},
  {Pace}, {Padilla}, {Pan}, {Pan}, {Parikh}, {Parker}, {Peirani}, {Pe{\~n}a
  Ram{\'\i}rez}, {Penny}, {Percival}, {Perez-Fournon}, {Pinsonneault},
  {Poidevin}, {Poovelil}, {Price-Whelan}, {B{\'a}rbara de Andrade Queiroz},
  {Raddick}, {Ray}, {Rembold}, {Riddle}, {Riffel}, {Riffel}, {Rix}, {Robin},
  {Rodr{\'\i}guez-Puebla}, {Roman-Lopes}, {Rom{\'a}n-Z{\'u}{\~n}iga}, {Rose},
  {Ross}, {Rossi}, {Rubin}, {Salvato}, {S{\'a}nchez}, {S{\'a}nchez-Gallego},
  {Sanderson}, {Santana Rojas}, {Sarceno}, {Sarmiento}, {Sayres}, {Sazonova},
  {Schaefer}, {Schiavon}, {Schlegel}, {Schneider}, {Schultheis}, {Schwope},
  {Serenelli}, {Serna}, {Shao}, {Shapiro}, {Sharma}, {Shen}, {Shetrone}, {Shu},
  {Simon}, {Skrutskie}, {Smethurst}, {Smith}, {Sobeck}, {Spoo}, {Sprague},
  {Stark}, {Stassun}, {Steinmetz}, {Stello}, {Stone-Martinez},
  {Storchi-Bergmann}, {Stringfellow}, {Stutz}, {Su}, {Taghizadeh-Popp},
  {Talbot}, {Tayar}, {Telles}, {Teske}, {Thakar}, {Theissen}, {Tkachenko},
  {Thomas}, {Tojeiro}, {Hernandez Toledo}, {Troup}, {Trump}, {Trussler},
  {Turner}, {Tuttle}, {Unda-Sanzana}, {V{\'a}zquez-Mata}, {Valentini},
  {Valenzuela}, {Vargas-Gonz{\'a}lez}, {Vargas-Maga{\~n}a}, {Alfaro},
  {Villanova}, {Vincenzo}, {Wake}, {Warfield}, {Washington}, {Weaver},
  {Weijmans}, {Weinberg}, {Weiss}, {Westfall}, {Wild}, {Wilde}, {Wilson},
  {Wilson}, {Wilson}, {Wolf}, {Wood-Vasey}, {Yan}, {Zamora}, {Zasowski},
  {Zhang}, {Zhao}, {Zheng}, {Zheng}, \& {Zhu}}]{2022ApJS..259...35A}
{Abdurro'uf}, {Accetta}, K., {Aerts}, C., {et~al.} 2022, \apjs, 259, 35

\bibitem[{{Ahumada} {et~al.}(2020){Ahumada}, {Allende Prieto}, {Almeida},
  {Anders}, {Anderson}, {Andrews}, {Anguiano}, {Arcodia}, {Armengaud},
  {Aubert}, {Avila}, {Avila-Reese}, {Badenes}, {Balland}, {Barger},
  {Barrera-Ballesteros}, {Basu}, {Bautista}, {Beaton}, {Beers}, {Benavides},
  {Bender}, {Bernardi}, {Bershady}, {Beutler}, {Bidin}, {Bird}, {Bizyaev},
  {Blanc}, {Blanton}, {Boquien}, {Borissova}, {Bovy}, {Brandt}, {Brinkmann},
  {Brownstein}, {Bundy}, {Bureau}, {Burgasser}, {Burtin}, {Cano-D{\'\i}az},
  {Capasso}, {Cappellari}, {Carrera}, {Chabanier}, {Chaplin}, {Chapman},
  {Cherinka}, {Chiappini}, {Doohyun Choi}, {Chojnowski}, {Chung}, {Clerc},
  {Coffey}, {Comerford}, {Comparat}, {da Costa}, {Cousinou}, {Covey}, {Crane},
  {Cunha}, {Ilha}, {Dai}, {Damsted}, {Darling}, {Davidson}, {Davies}, {Dawson},
  {De}, {de la Macorra}, {De Lee}, {Queiroz}, {Deconto Machado}, {de la Torre},
  {Dell'Agli}, {du Mas des Bourboux}, {Diamond-Stanic}, {Dillon}, {Donor},
  {Drory}, {Duckworth}, {Dwelly}, {Ebelke}, {Eftekharzadeh}, {Davis Eigenbrot},
  {Elsworth}, {Eracleous}, {Erfanianfar}, {Escoffier}, {Fan}, {Farr},
  {Fern{\'a}ndez-Trincado}, {Feuillet}, {Finoguenov}, {Fofie},
  {Fraser-McKelvie}, {Frinchaboy}, {Fromenteau}, {Fu}, {Galbany}, {Garcia},
  {Garc{\'\i}a-Hern{\'a}ndez}, {Garma Oehmichen}, {Ge}, {Geimba Maia},
  {Geisler}, {Gelfand}, {Goddy}, {Gonzalez-Perez}, {Grabowski}, {Green},
  {Grier}, {Guo}, {Guy}, {Harding}, {Hasselquist}, {Hawken}, {Hayes}, {Hearty},
  {Hekker}, {Hogg}, {Holtzman}, {Horta}, {Hou}, {Hsieh}, {Huber}, {Hunt}, {Ider
  Chitham}, {Imig}, {Jaber}, {Jimenez Angel}, {Johnson}, {Jones},
  {J{\"o}nsson}, {Jullo}, {Kim}, {Kinemuchi}, {Kirkpatrick}, {Kite}, {Klaene},
  {Kneib}, {Kollmeier}, {Kong}, {Kounkel}, {Krishnarao}, {Lacerna}, {Lan},
  {Lane}, {Law}, {Le Goff}, {Leung}, {Lewis}, {Li}, {Lian}, {Lin}, {Long},
  {Longa-Pe{\~n}a}, {Lundgren}, {Lyke}, {Mackereth}, {MacLeod}, {Majewski},
  {Manchado}, {Maraston}, {Martini}, {Masseron}, {Masters}, {Mathur},
  {McDermid}, {Merloni}, {Merrifield}, {M{\'e}sz{\'a}ros}, {Miglio}, {Minniti},
  {Minsley}, {Miyaji}, {Mohammad}, {Mosser}, {Mueller}, {Muna},
  {Mu{\~n}oz-Guti{\'e}rrez}, {Myers}, {Nadathur}, {Nair}, {Nandra}, {Correa do
  Nascimento}, {Nevin}, {Newman}, {Nidever}, {Nitschelm}, {Noterdaeme},
  {O'Connell}, {Olmstead}, {Oravetz}, {Oravetz}, {Osorio}, {Pace}, {Padilla},
  {Palanque-Delabrouille}, {Palicio}, {Pan}, {Pan}, {Parker}, {Paviot},
  {Peirani}, {Ram{\'r}ez}, {Penny}, {Percival}, {Perez-Fournon},
  {P{\'e}rez-R{\`a}fols}, {Petitjean}, {Pieri}, {Pinsonneault}, {Poovelil},
  {Povick}, {Prakash}, {Price-Whelan}, {Raddick}, {Raichoor}, {Ray}, {Rembold},
  {Rezaie}, {Riffel}, {Riffel}, {Rix}, {Robin}, {Roman-Lopes},
  {Rom{\'a}n-Z{\'u}{\~n}iga}, {Rose}, {Ross}, {Rossi}, {Rowlands}, {Rubin},
  {Salvato}, {S{\'a}nchez}, {S{\'a}nchez-Menguiano}, {S{\'a}nchez-Gallego},
  {Sayres}, {Schaefer}, {Schiavon}, {Schimoia}, {Schlafly}, {Schlegel},
  {Schneider}, {Schultheis}, {Schwope}, {Seo}, {Serenelli}, {Shafieloo},
  {Shamsi}, {Shao}, {Shen}, {Shetrone}, {Shirley}, {Silva Aguirre}, {Simon},
  {Skrutskie}, {Slosar}, {Smethurst}, {Sobeck}, {Sodi}, {Souto}, {Stark},
  {Stassun}, {Steinmetz}, {Stello}, {Stermer}, {Storchi-Bergmann},
  {Streblyanska}, {Stringfellow}, {Stutz}, {Su{\'a}rez}, {Sun},
  {Taghizadeh-Popp}, {Talbot}, {Tayar}, {Thakar}, {Theriault}, {Thomas},
  {Thomas}, {Tinker}, {Tojeiro}, {Toledo}, {Tremonti}, {Troup}, {Tuttle},
  {Unda-Sanzana}, {Valentini}, {Vargas-Gonz{\'a}lez}, {Vargas-Maga{\~n}a},
  {V{\'a}zquez-Mata}, {Vivek}, {Wake}, {Wang}, {Weaver}, {Weijmans}, {Wild},
  {Wilson}, {Wilson}, {Wolthuis}, {Wood-Vasey}, {Yan}, {Yang}, {Y{\`e}che},
  {Zamora}, {Zarrouk}, {Zasowski}, {Zhang}, {Zhao}, {Zhao}, {Zheng}, {Zheng},
  {Zhu}, \& {Zou}}]{2020ApJS..249....3A}
{Ahumada}, R., {Allende Prieto}, C., {Almeida}, A., {et~al.} 2020, \apjs, 249,
  3

\bibitem[{{Andrae} {et~al.}(2023){Andrae}, {Rix}, \&
  {Chandra}}]{2023arXiv230202611A}
{Andrae}, R., {Rix}, H.-W., \& {Chandra}, V. 2023, arXiv e-prints,
  arXiv:2302.02611

\bibitem[{{Aoki} {et~al.}(2022){Aoki}, {Li}, {Matsuno}, {Xing}, {Chen},
  {Christlieb}, {Honda}, {Ishigaki}, {Shi}, {Suda}, {Tominaga}, {Yan}, {Zhao},
  \& {Zhao}}]{2022ApJ...931..146A}
{Aoki}, W., {Li}, H., {Matsuno}, T., {et~al.} 2022, \apj, 931, 146

\bibitem[{{Arentsen} {et~al.}(2023){Arentsen}, {Aguado}, {Sestito},
  {Gonz{\'a}lez Hern{\'a}ndez}, {Martin}, {Starkenburg}, {Jablonka}, \&
  {Yuan}}]{2023MNRAS.519.5554A}
{Arentsen}, A., {Aguado}, D.~S., {Sestito}, F., {et~al.} 2023, \mnras, 519,
  5554

\bibitem[{{Bailer-Jones} {et~al.}(2021){Bailer-Jones}, {Rybizki}, {Fouesneau},
  {Demleitner}, \& {Andrae}}]{2021AJ....161..147B}
{Bailer-Jones}, C.~A.~L., {Rybizki}, J., {Fouesneau}, M., {Demleitner}, M., \&
  {Andrae}, R. 2021, \aj, 161, 147

\bibitem[{{Beers} {et~al.}(1985){Beers}, {Preston}, \&
  {Shectman}}]{1985AJ.....90.2089B}
{Beers}, T.~C., {Preston}, G.~W., \& {Shectman}, S.~A. 1985, \aj, 90, 2089

\bibitem[{{Beers} {et~al.}(1992){Beers}, {Preston}, \&
  {Shectman}}]{1992AJ....103.1987B}
---. 1992, \aj, 103, 1987

\bibitem[{{Bessell} {et~al.}(2011){Bessell}, {Bloxham}, {Schmidt}, {Keller},
  {Tisserand}, \& {Francis}}]{2011PASP..123..789B}
{Bessell}, M., {Bloxham}, G., {Schmidt}, B., {et~al.} 2011, \pasp, 123, 789

\bibitem[{{Bressan} {et~al.}(2012){Bressan}, {Marigo}, {Girardi}, {Salasnich},
  {Dal Cero}, {Rubele}, \& {Nanni}}]{2012MNRAS.427..127B}
{Bressan}, A., {Marigo}, P., {Girardi}, L., {et~al.} 2012, \mnras, 427, 127

\bibitem[{{Buder} {et~al.}(2021){Buder}, {Sharma}, {Kos}, {Amarsi},
  {Nordlander}, {Lind}, {Martell}, {Asplund}, {Bland-Hawthorn}, {Casey}, {de
  Silva}, {D'Orazi}, {Freeman}, {Hayden}, {Lewis}, {Lin}, {Schlesinger},
  {Simpson}, {Stello}, {Zucker}, {Zwitter}, {Beeson}, {Buck}, {Casagrande},
  {Clark}, {{\v{C}}otar}, {da Costa}, {de Grijs}, {Feuillet}, {Horner},
  {Kafle}, {Khanna}, {Kobayashi}, {Liu}, {Montet}, {Nandakumar}, {Nataf},
  {Ness}, {Spina}, {Tepper-Garc{\'\i}a}, {Ting}, {Traven},
  {Vogrin{\v{c}}i{\v{c}}}, {Wittenmyer}, {Wyse}, {{\v{Z}}erjal}, \& {Galah
  Collaboration}}]{2021MNRAS.506..150B}
{Buder}, S., {Sharma}, S., {Kos}, J., {et~al.} 2021, \mnras, 506, 150

\bibitem[{{Cantat-Gaudin} {et~al.}(2023){Cantat-Gaudin}, {Fouesneau}, {Rix},
  {Brown}, {Castro-Ginard}, {Kostrzewa-Rutkowska}, {Drimmel}, {Hogg}, {Casey},
  {Khanna}, {Oh}, {Price-Whelan}, {Belokurov}, {Saydjari}, \&
  {Green}}]{2023A&A...669A..55C}
{Cantat-Gaudin}, T., {Fouesneau}, M., {Rix}, H.-W., {et~al.} 2023, \aap, 669,
  A55

\bibitem[{{Casagrande} {et~al.}(2019){Casagrande}, {Wolf}, {Mackey},
  {Nordlander}, {Yong}, \& {Bessell}}]{2019MNRAS.482.2770C}
{Casagrande}, L., {Wolf}, C., {Mackey}, A.~D., {et~al.} 2019, \mnras, 482, 2770

\bibitem[{{Castro-Ginard} {et~al.}(2023){Castro-Ginard}, {Brown},
  {Kostrzewa-Rutkowska}, {Cantat-Gaudin}, {Drimmel}, {Oh}, {Belokurov},
  {Casey}, {Fouesneau}, {Khanna}, {Price-Whelan}, \&
  {Rix}}]{2023arXiv230317738C}
{Castro-Ginard}, A., {Brown}, A.~G.~A., {Kostrzewa-Rutkowska}, Z., {et~al.}
  2023, arXiv e-prints, arXiv:2303.17738

\bibitem[{{Cenarro} {et~al.}(2019){Cenarro}, {Moles},
  {Crist{\'o}bal-Hornillos}, {Mar{\'\i}n-Franch}, {Ederoclite}, {Varela},
  {L{\'o}pez-Sanjuan}, {Hern{\'a}ndez-Monteagudo}, {Angulo}, {V{\'a}zquez
  Rami{\'o}}, {Viironen}, {Bonoli}, {Orsi}, {Hurier}, {San Roman}, {Greisel},
  {Vilella-Rojo}, {D{\'\i}az-Garc{\'\i}a}, {Logro{\~n}o-Garc{\'\i}a},
  {Gurung-L{\'o}pez}, {Spinoso}, {Izquierdo-Villalba}, {Aguerri}, {Allende
  Prieto}, {Bonatto}, {Carvano}, {Chies-Santos}, {Daflon}, {Dupke},
  {Falc{\'o}n-Barroso}, {Gon{\c{c}}alves}, {Jim{\'e}nez-Teja}, {Molino},
  {Placco}, {Solano}, {Whitten}, {Abril}, {Ant{\'o}n}, {Bello}, {Bielsa de
  Toledo}, {Castillo-Ram{\'\i}rez}, {Chueca}, {Civera},
  {D{\'\i}az-Mart{\'\i}n}, {Dom{\'\i}nguez-Mart{\'\i}nez},
  {Garzar{\'a}n-Calderaro}, {Hern{\'a}ndez-Fuertes}, {Iglesias-Marzoa},
  {I{\~n}iguez}, {Jim{\'e}nez Ruiz}, {Kruuse}, {Lamadrid}, {Lasso-Cabrera},
  {L{\'o}pez-Alegre}, {L{\'o}pez-Sainz}, {Ma{\'\i}cas}, {Moreno-Signes},
  {Muniesa}, {Rodr{\'\i}guez-Llano}, {Rueda-Teruel}, {Rueda-Teruel},
  {Soriano-Lagu{\'\i}a}, {Tilve}, {Valdivielso}, {Yanes-D{\'\i}az}, {Alcaniz},
  {Mendes de Oliveira}, {Sodr{\'e}}, {Coelho}, {Lopes de Oliveira}, {Tamm},
  {Xavier}, {Abramo}, {Akras}, {Alfaro}, {Alvarez-Candal}, {Ascaso}, {Beasley},
  {Beers}, {Borges Fernandes}, {Bruzual}, {Buzzo}, {Carrasco}, {Cepa},
  {Cortesi}, {Costa-Duarte}, {De Pr{\'a}}, {Favole}, {Galarza}, {Galbany},
  {Garcia}, {Gonz{\'a}lez Delgado}, {Gonz{\'a}lez-Serrano},
  {Guti{\'e}rrez-Soto}, {Hernandez-Jimenez}, {Kanaan}, {Kuncarayakti},
  {Landim}, {Laur}, {Licandro}, {Lima Neto}, {Lyman}, {Ma{\'\i}z
  Apell{\'a}niz}, {Miralda-Escud{\'e}}, {Morate}, {Nogueira-Cavalcante},
  {Novais}, {Oncins}, {Oteo}, {Overzier}, {Pereira}, {Rebassa-Mansergas},
  {Reis}, {Roig}, {Sako}, {Salvador-Rusi{\~n}ol}, {Sampedro},
  {S{\'a}nchez-Bl{\'a}zquez}, {Santos}, {Schmidtobreick}, {Siffert}, {Telles},
  \& {Vilchez}}]{2019A&A...622A.176C}
{Cenarro}, A.~J., {Moles}, M., {Crist{\'o}bal-Hornillos}, D., {et~al.} 2019,
  \aap, 622, A176

\bibitem[{{Chambers} {et~al.}(2016){Chambers}, {Magnier}, {Metcalfe},
  {Flewelling}, {Huber}, {Waters}, {Denneau}, {Draper}, {Farrow}, {Finkbeiner},
  {Holmberg}, {Koppenhoefer}, {Price}, {Rest}, {Saglia}, {Schlafly}, {Smartt},
  {Sweeney}, {Wainscoat}, {Burgett}, {Chastel}, {Grav}, {Heasley}, {Hodapp},
  {Jedicke}, {Kaiser}, {Kudritzki}, {Luppino}, {Lupton}, {Monet}, {Morgan},
  {Onaka}, {Shiao}, {Stubbs}, {Tonry}, {White}, {Ba{\~n}ados}, {Bell},
  {Bender}, {Bernard}, {Boegner}, {Boffi}, {Botticella}, {Calamida},
  {Casertano}, {Chen}, {Chen}, {Cole}, {Deacon}, {Frenk}, {Fitzsimmons},
  {Gezari}, {Gibbs}, {Goessl}, {Goggia}, {Gourgue}, {Goldman}, {Grant},
  {Grebel}, {Hambly}, {Hasinger}, {Heavens}, {Heckman}, {Henderson}, {Henning},
  {Holman}, {Hopp}, {Ip}, {Isani}, {Jackson}, {Keyes}, {Koekemoer}, {Kotak},
  {Le}, {Liska}, {Long}, {Lucey}, {Liu}, {Martin}, {Masci}, {McLean}, {Mindel},
  {Misra}, {Morganson}, {Murphy}, {Obaika}, {Narayan}, {Nieto-Santisteban},
  {Norberg}, {Peacock}, {Pier}, {Postman}, {Primak}, {Rae}, {Rai}, {Riess},
  {Riffeser}, {Rix}, {R{\"o}ser}, {Russel}, {Rutz}, {Schilbach}, {Schultz},
  {Scolnic}, {Strolger}, {Szalay}, {Seitz}, {Small}, {Smith}, {Soderblom},
  {Taylor}, {Thomson}, {Taylor}, {Thakar}, {Thiel}, {Thilker}, {Unger},
  {Urata}, {Valenti}, {Wagner}, {Walder}, {Walter}, {Watters}, {Werner},
  {Wood-Vasey}, \& {Wyse}}]{2016arXiv161205560C}
{Chambers}, K.~C., {Magnier}, E.~A., {Metcalfe}, N., {et~al.} 2016, arXiv
  e-prints, arXiv:1612.05560

\bibitem[{{Chiti} {et~al.}(2021){Chiti}, {Frebel}, {Mardini}, {Daniel}, {Ou},
  \& {Uvarova}}]{2021ApJS..254...31C}
{Chiti}, A., {Frebel}, A., {Mardini}, M.~K., {et~al.} 2021, \apjs, 254, 31

\bibitem[{{Christlieb}(2003)}]{2003RvMA...16..191C}
{Christlieb}, N. 2003, Reviews in Modern Astronomy, 16, 191

\bibitem[{{De Silva} {et~al.}(2015){De Silva}, {Freeman}, {Bland-Hawthorn},
  {Martell}, {de Boer}, {Asplund}, {Keller}, {Sharma}, {Zucker}, {Zwitter},
  {Anguiano}, {Bacigalupo}, {Bayliss}, {Beavis}, {Bergemann}, {Campbell},
  {Cannon}, {Carollo}, {Casagrande}, {Casey}, {Da Costa}, {D'Orazi}, {Dotter},
  {Duong}, {Heger}, {Ireland}, {Kafle}, {Kos}, {Lattanzio}, {Lewis}, {Lin},
  {Lind}, {Munari}, {Nataf}, {O'Toole}, {Parker}, {Reid}, {Schlesinger},
  {Sheinis}, {Simpson}, {Stello}, {Ting}, {Traven}, {Watson}, {Wittenmyer},
  {Yong}, \& {{\v{Z}}erjal}}]{2015MNRAS.449.2604D}
{De Silva}, G.~M., {Freeman}, K.~C., {Bland-Hawthorn}, J., {et~al.} 2015,
  \mnras, 449, 2604

\bibitem[{{Deng} {et~al.}(2012){Deng}, {Newberg}, {Liu}, {Carlin}, {Beers},
  {Chen}, {Chen}, {Christlieb}, {Grillmair}, {Guhathakurta}, {Han}, {Hou},
  {Lee}, {L{\'e}pine}, {Li}, {Liu}, {Pan}, {Sellwood}, {Wang}, {Wang}, {Yang},
  {Yanny}, {Zhang}, {Zhang}, {Zheng}, \& {Zhu}}]{2012RAA....12..735D}
{Deng}, L.-C., {Newberg}, H.~J., {Liu}, C., {et~al.} 2012, Research in
  Astronomy and Astrophysics, 12, 735

\bibitem[{{Ezzeddine} {et~al.}(2020){Ezzeddine}, {Rasmussen}, {Frebel},
  {Chiti}, {Hinojisa}, {Placco}, {Ji}, {Beers}, {Hansen}, {Roederer}, {Sakari},
  \& {Melendez}}]{2020ApJ...898..150E}
{Ezzeddine}, R., {Rasmussen}, K., {Frebel}, A., {et~al.} 2020, \apj, 898, 150

\bibitem[{{Fan} {et~al.}(2023){Fan}, {Zhao}, {Wang}, {Zheng}, {Zhao}, {Li},
  {Chen}, {Yuan}, {Li}, {Tan}, {Song}, {Zuo}, {Huang}, {Luo}, {Esamdin}, {Ma},
  {Li}, {Song}, {Grupp}, {Zhao}, {Ehgamberdiev}, {Burkhonov}, {Feng}, {Bai},
  {Zhang}, {Niu}, {Khodjaev}, {Khafizov}, {Asfandiyarov}, {Shaymanov},
  {Karimov}, {Yuldashev}, {Lu}, {Zhaori}, {Hong}, {Hu}, {Liu}, \&
  {Xu}}]{2023arXiv230615611F}
{Fan}, Z., {Zhao}, G., {Wang}, W., {et~al.} 2023, arXiv e-prints,
  arXiv:2306.15611

\bibitem[{{Gaia Collaboration} {et~al.}(2021){Gaia Collaboration}, {Brown},
  {Vallenari}, {Prusti}, {de Bruijne}, {Babusiaux}, {Biermann}, {Creevey},
  {Evans}, {Eyer}, {Hutton}, {Jansen}, {Jordi}, {Klioner}, {Lammers},
  {Lindegren}, {Luri}, {Mignard}, {Panem}, {Pourbaix}, {Randich}, {Sartoretti},
  {Soubiran}, {Walton}, {Arenou}, {Bailer-Jones}, {Bastian}, {Cropper},
  {Drimmel}, {Katz}, {Lattanzi}, {van Leeuwen}, {Bakker}, {Cacciari},
  {Casta{\~n}eda}, {De Angeli}, {Ducourant}, {Fabricius}, {Fouesneau},
  {Fr{\'e}mat}, {Guerra}, {Guerrier}, {Guiraud}, {Jean-Antoine Piccolo},
  {Masana}, {Messineo}, {Mowlavi}, {Nicolas}, {Nienartowicz}, {Pailler},
  {Panuzzo}, {Riclet}, {Roux}, {Seabroke}, {Sordo}, {Tanga}, {Th{\'e}venin},
  {Gracia-Abril}, {Portell}, {Teyssier}, {Altmann}, {Andrae}, {Bellas-Velidis},
  {Benson}, {Berthier}, {Blomme}, {Brugaletta}, {Burgess}, {Busso}, {Carry},
  {Cellino}, {Cheek}, {Clementini}, {Damerdji}, {Davidson}, {Delchambre},
  {Dell'Oro}, {Fern{\'a}ndez-Hern{\'a}ndez}, {Galluccio}, {Garc{\'\i}a-Lario},
  {Garcia-Reinaldos}, {Gonz{\'a}lez-N{\'u}{\~n}ez}, {Gosset}, {Haigron},
  {Halbwachs}, {Hambly}, {Harrison}, {Hatzidimitriou}, {Heiter},
  {Hern{\'a}ndez}, {Hestroffer}, {Hodgkin}, {Holl}, {Jan{\ss}en}, {Jevardat de
  Fombelle}, {Jordan}, {Krone-Martins}, {Lanzafame}, {L{\"o}ffler}, {Lorca},
  {Manteiga}, {Marchal}, {Marrese}, {Moitinho}, {Mora}, {Muinonen}, {Osborne},
  {Pancino}, {Pauwels}, {Petit}, {Recio-Blanco}, {Richards}, {Riello},
  {Rimoldini}, {Robin}, {Roegiers}, {Rybizki}, {Sarro}, {Siopis}, {Smith},
  {Sozzetti}, {Ulla}, {Utrilla}, {van Leeuwen}, {van Reeven}, {Abbas}, {Abreu
  Aramburu}, {Accart}, {Aerts}, {Aguado}, {Ajaj}, {Altavilla}, {{\'A}lvarez},
  {{\'A}lvarez Cid-Fuentes}, {Alves}, {Anderson}, {Anglada Varela}, {Antoja},
  {Audard}, {Baines}, {Baker}, {Balaguer-N{\'u}{\~n}ez}, {Balbinot}, {Balog},
  {Barache}, {Barbato}, {Barros}, {Barstow}, {Bartolom{\'e}}, {Bassilana},
  {Bauchet}, {Baudesson-Stella}, {Becciani}, {Bellazzini}, {Bernet}, {Bertone},
  {Bianchi}, {Blanco-Cuaresma}, {Boch}, {Bombrun}, {Bossini}, {Bouquillon},
  {Bragaglia}, {Bramante}, {Breedt}, {Bressan}, {Brouillet}, {Bucciarelli},
  {Burlacu}, {Busonero}, {Butkevich}, {Buzzi}, {Caffau}, {Cancelliere},
  {C{\'a}novas}, {Cantat-Gaudin}, {Carballo}, {Carlucci}, {Carnerero},
  {Carrasco}, {Casamiquela}, {Castellani}, {Castro-Ginard}, {Castro Sampol},
  {Chaoul}, {Charlot}, {Chemin}, {Chiavassa}, {Cioni}, {Comoretto}, {Cooper},
  {Cornez}, {Cowell}, {Crifo}, {Crosta}, {Crowley}, {Dafonte}, {Dapergolas},
  {David}, {David}, {de Laverny}, {De Luise}, {De March}, {De Ridder}, {de
  Souza}, {de Teodoro}, {de Torres}, {del Peloso}, {del Pozo}, {Delbo},
  {Delgado}, {Delgado}, {Delisle}, {Di Matteo}, {Diakite}, {Diener},
  {Distefano}, {Dolding}, {Eappachen}, {Edvardsson}, {Enke}, {Esquej}, {Fabre},
  {Fabrizio}, {Faigler}, {Fedorets}, {Fernique}, {Fienga}, {Figueras},
  {Fouron}, {Fragkoudi}, {Fraile}, {Franke}, {Gai}, {Garabato},
  {Garcia-Gutierrez}, {Garc{\'\i}a-Torres}, {Garofalo}, {Gavras}, {Gerlach},
  {Geyer}, {Giacobbe}, {Gilmore}, {Girona}, {Giuffrida}, {Gomel}, {Gomez},
  {Gonzalez-Santamaria}, {Gonz{\'a}lez-Vidal}, {Granvik},
  {Guti{\'e}rrez-S{\'a}nchez}, {Guy}, {Hauser}, {Haywood}, {Helmi}, {Hidalgo},
  {Hilger}, {H{\l}adczuk}, {Hobbs}, {Holland}, {Huckle}, {Jasniewicz},
  {Jonker}, {Juaristi Campillo}, {Julbe}, {Karbevska}, {Kervella}, {Khanna},
  {Kochoska}, {Kontizas}, {Kordopatis}, {Korn}, {Kostrzewa-Rutkowska},
  {Kruszy{\'n}ska}, {Lambert}, {Lanza}, {Lasne}, {Le Campion}, {Le Fustec},
  {Lebreton}, {Lebzelter}, {Leccia}, {Leclerc}, {Lecoeur-Taibi}, {Liao},
  {Licata}, {Lindstr{\o}m}, {Lister}, {Livanou}, {Lobel}, {Madrero Pardo},
  {Managau}, {Mann}, {Marchant}, {Marconi}, {Marcos Santos}, {Marinoni},
  {Marocco}, {Marshall}, {Martin Polo}, {Mart{\'\i}n-Fleitas}, {Masip},
  {Massari}, {Mastrobuono-Battisti}, {Mazeh}, {McMillan}, {Messina},
  {Michalik}, {Millar}, {Mints}, {Molina}, {Molinaro}, {Moln{\'a}r},
  {Montegriffo}, {Mor}, {Morbidelli}, {Morel}, {Morris}, {Mulone}, {Munoz},
  {Muraveva}, {Murphy}, {Musella}, {Noval}, {Ord{\'e}novic}, {Orr{\`u}},
  {Osinde}, {Pagani}, {Pagano}, {Palaversa}, {Palicio}, {Panahi}, {Pawlak},
  {Pe{\~n}alosa Esteller}, {Penttil{\"a}}, {Piersimoni}, {Pineau}, {Plachy},
  {Plum}, {Poggio}, {Poretti}, {Poujoulet}, {Pr{\v{s}}a}, {Pulone}, {Racero},
  {Ragaini}, {Rainer}, {Raiteri}, {Rambaux}, {Ramos}, {Ramos-Lerate}, {Re
  Fiorentin}, {Regibo}, {Reyl{\'e}}, {Ripepi}, {Riva}, {Rixon}, {Robichon},
  {Robin}, {Roelens}, {Rohrbasser}, {Romero-G{\'o}mez}, {Rowell}, {Royer},
  {Rybicki}, {Sadowski}, {Sagrist{\`a} Sell{\'e}s}, {Sahlmann}, {Salgado},
  {Salguero}, {Samaras}, {Sanchez Gimenez}, {Sanna}, {Santove{\~n}a},
  {Sarasso}, {Schultheis}, {Sciacca}, {Segol}, {Segovia}, {S{\'e}gransan},
  {Semeux}, {Shahaf}, {Siddiqui}, {Siebert}, {Siltala}, {Slezak}, {Smart},
  {Solano}, {Solitro}, {Souami}, {Souchay}, {Spagna}, {Spoto}, {Steele},
  {Steidelm{\"u}ller}, {Stephenson}, {S{\"u}veges}, {Szabados}, {Szegedi-Elek},
  {Taris}, {Tauran}, {Taylor}, {Teixeira}, {Thuillot}, {Tonello}, {Torra},
  {Torra}, {Turon}, {Unger}, {Vaillant}, {van Dillen}, {Vanel}, {Vecchiato},
  {Viala}, {Vicente}, {Voutsinas}, {Weiler}, {Wevers}, {Wyrzykowski}, {Yoldas},
  {Yvard}, {Zhao}, {Zorec}, {Zucker}, {Zurbach}, \&
  {Zwitter}}]{2021A&A...649A...1G}
{Gaia Collaboration}, {Brown}, A.~G.~A., {Vallenari}, A., {et~al.} 2021, \aap,
  649, A1

\bibitem[{{Gaia Collaboration} {et~al.}(2022){Gaia Collaboration},
  {Montegriffo}, {Bellazzini}, {De Angeli}, {Andrae}, {Barstow}, {Bossini},
  {Bragaglia}, {Burgess}, {Cacciari}, {Carrasco}, {Chornay}, {Delchambre},
  {Evans}, {Fouesneau}, {Fremat}, {Garabato}, {Jordi}, {Manteiga}, {Massari},
  {Palaversa}, {Pancino}, {Riello}, {Ruz Mieres}, {Sanna}, {Santovena},
  {Sordo}, {Vallenari}, {Walton}, \& {DPAC}}]{2022arXiv220606215G}
{Gaia Collaboration}, {Montegriffo}, P., {Bellazzini}, M., {et~al.} 2022, arXiv
  e-prints, arXiv:2206.06215

\bibitem[{{Green} {et~al.}(2019){Green}, {Schlafly}, {Zucker}, {Speagle}, \&
  {Finkbeiner}}]{2019ApJ...887...93G}
{Green}, G.~M., {Schlafly}, E., {Zucker}, C., {Speagle}, J.~S., \&
  {Finkbeiner}, D. 2019, \apj, 887, 93

\bibitem[{{Hansen} {et~al.}(2018){Hansen}, {Holmbeck}, {Beers}, {Placco},
  {Roederer}, {Frebel}, {Sakari}, {Simon}, \& {Thompson}}]{2018ApJ...858...92H}
{Hansen}, T.~T., {Holmbeck}, E.~M., {Beers}, T.~C., {et~al.} 2018, \apj, 858,
  92

\bibitem[{{Holmbeck} {et~al.}(2020){Holmbeck}, {Hansen}, {Beers}, {Placco},
  {Whitten}, {Rasmussen}, {Roederer}, {Ezzeddine}, {Sakari}, {Frebel}, {Drout},
  {Simon}, {Thompson}, {Bland-Hawthorn}, {Gibson}, {Grebel}, {Kordopatis},
  {Kunder}, {Mel{\'e}ndez}, {Navarro}, {Reid}, {Seabroke}, {Steinmetz},
  {Watson}, \& {Wyse}}]{2020ApJS..249...30H}
{Holmbeck}, E.~M., {Hansen}, T.~T., {Beers}, T.~C., {et~al.} 2020, \apjs, 249,
  30

\bibitem[{{Huang} {et~al.}(2018){Huang}, {Liu}, {Chen}, {Zhang}, {Yuan},
  {Xiang}, {Wang}, \& {Tian}}]{2018AJ....156...90H}
{Huang}, Y., {Liu}, X.~W., {Chen}, B.~Q., {et~al.} 2018, \aj, 156, 90

\bibitem[{{Huang} {et~al.}(2019){Huang}, {Chen}, {Yuan}, {Zhang}, {Xiang},
  {Wang}, {Wang}, {Wolf}, {Liu}, \& {Liu}}]{2019ApJS..243....7H}
{Huang}, Y., {Chen}, B.~Q., {Yuan}, H.~B., {et~al.} 2019, \apjs, 243, 7

\bibitem[{{Huang} {et~al.}(2022){Huang}, {Beers}, {Wolf}, {Lee}, {Onken},
  {Yuan}, {Shank}, {Zhang}, {Wang}, {Shi}, \& {Fan}}]{2022ApJ...925..164H}
{Huang}, Y., {Beers}, T.~C., {Wolf}, C., {et~al.} 2022, \apj, 925, 164

\bibitem[{{Katz} {et~al.}(2022){Katz}, {Sartoretti}, {Guerrier}, {Panuzzo},
  {Seabroke}, {Th{\'e}venin}, {Cropper}, {Benson}, {Blomme}, {Haigron},
  {Marchal}, {Smith}, {Baker}, {Chemin}, {Damerdji}, {David}, {Dolding},
  {Fr{\'e}mat}, {Gosset}, {Jan{\ss}en}, {Jasniewicz}, {Lobel}, {Plum},
  {Samaras}, {Snaith}, {Soubiran}, {Vanel}, {Zwitter}, {Antoja}, {Arenou},
  {Babusiaux}, {Brouillet}, {Caffau}, {Di Matteo}, {Fabre}, {Fabricius},
  {Frakgoudi}, {Haywood}, {Huckle}, {Hottier}, {Lasne}, {Leclerc},
  {Mastrobuono-Battisti}, {Royer}, {Teyssier}, {Zorec}, {Crifo}, {Jean-Antoine
  Piccolo}, {Turon}, \& {Viala}}]{2022arXiv220605902K}
{Katz}, D., {Sartoretti}, P., {Guerrier}, A., {et~al.} 2022, arXiv e-prints,
  arXiv:2206.05902

\bibitem[{{Kunder} {et~al.}(2017){Kunder}, {Kordopatis}, {Steinmetz},
  {Zwitter}, {McMillan}, {Casagrande}, {Enke}, {Wojno}, {Valentini},
  {Chiappini}, {Matijevi{\v{c}}}, {Siviero}, {de Laverny}, {Recio-Blanco},
  {Bijaoui}, {Wyse}, {Binney}, {Grebel}, {Helmi}, {Jofre}, {Antoja}, {Gilmore},
  {Siebert}, {Famaey}, {Bienaym{\'e}}, {Gibson}, {Freeman}, {Navarro},
  {Munari}, {Seabroke}, {Anguiano}, {{\v{Z}}erjal}, {Minchev}, {Reid},
  {Bland-Hawthorn}, {Kos}, {Sharma}, {Watson}, {Parker}, {Scholz}, {Burton},
  {Cass}, {Hartley}, {Fiegert}, {Stupar}, {Ritter}, {Hawkins}, {Gerhard},
  {Chaplin}, {Davies}, {Elsworth}, {Lund}, {Miglio}, \&
  {Mosser}}]{2017AJ....153...75K}
{Kunder}, A., {Kordopatis}, G., {Steinmetz}, M., {et~al.} 2017, \aj, 153, 75

\bibitem[{{Li} {et~al.}(2022){Li}, {Aoki}, {Matsuno}, {Xing}, {Suda},
  {Tominaga}, {Chen}, {Honda}, {Ishigaki}, {Shi}, {Zhao}, \&
  {Zhao}}]{2022ApJ...931..147L}
{Li}, H., {Aoki}, W., {Matsuno}, T., {et~al.} 2022, \apj, 931, 147

\bibitem[{{Liu} {et~al.}(2014){Liu}, {Yuan}, {Huo}, {Deng}, {Hou}, {Zhao},
  {Zhao}, {Shi}, {Luo}, {Xiang}, {Zhang}, {Huang}, \&
  {Zhang}}]{2014IAUS..298..310L}
{Liu}, X.~W., {Yuan}, H.~B., {Huo}, Z.~Y., {et~al.} 2014, in Setting the scene
  for Gaia and LAMOST, ed. S.~{Feltzing}, G.~{Zhao}, N.~A. {Walton}, \&
  P.~{Whitelock}, Vol. 298, 310--321

\bibitem[{{Lucchesi} {et~al.}(2022){Lucchesi}, {Lardo}, {Jablonka}, {Sestito},
  {Mashonkina}, {Arentsen}, {Suter}, {Venn}, {Martin}, {Starkenburg}, {Aguado},
  {Hill}, {Kordopatis}, {Navarro}, {Gonz{\'a}lez Hern{\'a}ndez}, {Malhan}, \&
  {Yuan}}]{2022MNRAS.511.1004L}
{Lucchesi}, R., {Lardo}, C., {Jablonka}, P., {et~al.} 2022, \mnras, 511, 1004

\bibitem[{{Majewski} {et~al.}(2017){Majewski}, {Schiavon}, {Frinchaboy},
  {Allende Prieto}, {Barkhouser}, {Bizyaev}, {Blank}, {Brunner}, {Burton},
  {Carrera}, {Chojnowski}, {Cunha}, {Epstein}, {Fitzgerald}, {Garc{\'\i}a
  P{\'e}rez}, {Hearty}, {Henderson}, {Holtzman}, {Johnson}, {Lam}, {Lawler},
  {Maseman}, {M{\'e}sz{\'a}ros}, {Nelson}, {Nguyen}, {Nidever}, {Pinsonneault},
  {Shetrone}, {Smee}, {Smith}, {Stolberg}, {Skrutskie}, {Walker}, {Wilson},
  {Zasowski}, {Anders}, {Basu}, {Beland}, {Blanton}, {Bovy}, {Brownstein},
  {Carlberg}, {Chaplin}, {Chiappini}, {Eisenstein}, {Elsworth}, {Feuillet},
  {Fleming}, {Galbraith-Frew}, {Garc{\'\i}a}, {Garc{\'\i}a-Hern{\'a}ndez},
  {Gillespie}, {Girardi}, {Gunn}, {Hasselquist}, {Hayden}, {Hekker}, {Ivans},
  {Kinemuchi}, {Klaene}, {Mahadevan}, {Mathur}, {Mosser}, {Muna}, {Munn},
  {Nichol}, {O'Connell}, {Parejko}, {Robin}, {Rocha-Pinto}, {Schultheis},
  {Serenelli}, {Shane}, {Silva Aguirre}, {Sobeck}, {Thompson}, {Troup},
  {Weinberg}, \& {Zamora}}]{2017AJ....154...94M}
{Majewski}, S.~R., {Schiavon}, R.~P., {Frinchaboy}, P.~M., {et~al.} 2017, \aj,
  154, 94

\bibitem[{{Marigo} {et~al.}(2017){Marigo}, {Girardi}, {Bressan}, {Rosenfield},
  {Aringer}, {Chen}, {Dussin}, {Nanni}, {Pastorelli}, {Rodrigues}, {Trabucchi},
  {Bladh}, {Dalcanton}, {Groenewegen}, {Montalb{\'a}n}, \&
  {Wood}}]{2017ApJ...835...77M}
{Marigo}, P., {Girardi}, L., {Bressan}, A., {et~al.} 2017, \apj, 835, 77

\bibitem[{{Mendes de Oliveira} {et~al.}(2019){Mendes de Oliveira}, {Ribeiro},
  {Schoenell}, {Kanaan}, {Overzier}, {Molino}, {Sampedro}, {Coelho}, {Barbosa},
  {Cortesi}, {Costa-Duarte}, {Herpich}, {Hernandez-Jimenez}, {Placco},
  {Xavier}, {Abramo}, {Saito}, {Chies-Santos}, {Ederoclite}, {Lopes de
  Oliveira}, {Gon{\c{c}}alves}, {Akras}, {Almeida}, {Almeida-Fernandes},
  {Beers}, {Bonatto}, {Bonoli}, {Cypriano}, {Vinicius-Lima}, {de Souza},
  {Fabiano de Souza}, {Ferrari}, {Gon{\c{c}}alves}, {Gonzalez},
  {Guti{\'e}rrez-Soto}, {Hartmann}, {Jaffe}, {Kerber}, {Lima-Dias}, {Lopes},
  {Menendez-Delmestre}, {Nakazono}, {Novais}, {Ortega-Minakata}, {Pereira},
  {Perottoni}, {Queiroz}, {Reis}, {Santos}, {Santos-Silva}, {Santucci},
  {Barbosa}, {Siffert}, {Sodr{\'e}}, {Torres-Flores}, {Westera}, {Whitten},
  {Alcaniz}, {Alonso-Garc{\'\i}a}, {Alencar}, {Alvarez-Candal}, {Amram},
  {Azanha}, {Barb{\'a}}, {Bernardinelli}, {Borges Fernandes}, {Branco},
  {Brito-Silva}, {Buzzo}, {Caffer}, {Campillay}, {Cano}, {Carvano}, {Castejon},
  {Cid Fernandes}, {Dantas}, {Daflon}, {Damke}, {de la Reza}, {de Melo de
  Azevedo}, {De Paula}, {Diem}, {Donnerstein}, {Dors}, {Dupke}, {Eikenberry},
  {Escudero}, {Faifer}, {Far{\'\i}as}, {Fernandes}, {Fernandes}, {Fontes},
  {Galarza}, {Hirata}, {Katena}, {Gregorio-Hetem},
  {Hern{\'a}ndez-Fern{\'a}ndez}, {Izzo}, {Jaque Arancibia}, {Jatenco-Pereira},
  {Jim{\'e}nez-Teja}, {Kann}, {Krabbe}, {Labayru}, {Lazzaro}, {Lima Neto},
  {Lopes}, {Magalh{\~a}es}, {Makler}, {de Menezes}, {Miralda-Escud{\'e}},
  {Monteiro-Oliveira}, {Montero-Dorta}, {Mu{\~n}oz-Elgueta}, {Nemmen}, {Nilo
  Castell{\'o}n}, {Oliveira}, {Ort{\'\i}z}, {Pattaro}, {Pereira}, {Quint},
  {Riguccini}, {Rocha Pinto}, {Rodrigues}, {Roig}, {Rossi}, {Saha}, {Santos},
  {Schnorr M{\"u}ller}, {Sesto}, {Silva}, {Smith Castelli}, {Teixeira},
  {Telles}, {Thom de Souza}, {Th{\"o}ne}, {Trevisan}, {de Ugarte Postigo},
  {Urrutia-Viscarra}, {Veiga}, {Vika}, {Vitorelli}, {Werle}, {Werner}, \&
  {Zaritsky}}]{2019MNRAS.489..241M}
{Mendes de Oliveira}, C., {Ribeiro}, T., {Schoenell}, W., {et~al.} 2019,
  \mnras, 489, 241

\bibitem[{{Nordstr{\"o}m} {et~al.}(2004){Nordstr{\"o}m}, {Mayor}, {Andersen},
  {Holmberg}, {Pont}, {J{\o}rgensen}, {Olsen}, {Udry}, \&
  {Mowlavi}}]{2004A&A...418..989N}
{Nordstr{\"o}m}, B., {Mayor}, M., {Andersen}, J., {et~al.} 2004, \aap, 418, 989

\bibitem[{{Norris} \& {Yong}(2019)}]{2019ApJ...879...37N}
{Norris}, J.~E., \& {Yong}, D. 2019, \apj, 879, 37

\bibitem[{{Onken} {et~al.}(2022){Onken}, {Wolf}, {Bian}, {Fan}, {Hon},
  {Raithel}, {Tisserand}, \& {Lai}}]{2022MNRAS.511..572O}
{Onken}, C.~A., {Wolf}, C., {Bian}, F., {et~al.} 2022, \mnras, 511, 572

\bibitem[{{Onken} {et~al.}(2019){Onken}, {Wolf}, {Bessell}, {Chang}, {Da
  Costa}, {Luvaul}, {Mackey}, {Schmidt}, \& {Shao}}]{2019PASA...36...33O}
{Onken}, C.~A., {Wolf}, C., {Bessell}, M.~S., {et~al.} 2019, \pasa, 36, e033

\bibitem[{{Riello} {et~al.}(2021){Riello}, {De Angeli}, {Evans}, {Montegriffo},
  {Carrasco}, {Busso}, {Palaversa}, {Burgess}, {Diener}, {Davidson}, {Rowell},
  {Fabricius}, {Jordi}, {Bellazzini}, {Pancino}, {Harrison}, {Cacciari}, {van
  Leeuwen}, {Hambly}, {Hodgkin}, {Osborne}, {Altavilla}, {Barstow}, {Brown},
  {Castellani}, {Cowell}, {De Luise}, {Gilmore}, {Giuffrida}, {Hidalgo},
  {Holland}, {Marinoni}, {Pagani}, {Piersimoni}, {Pulone}, {Ragaini}, {Rainer},
  {Richards}, {Sanna}, {Walton}, {Weiler}, \& {Yoldas}}]{2021A&A...649A...3R}
{Riello}, M., {De Angeli}, F., {Evans}, D.~W., {et~al.} 2021, \aap, 649, A3

\bibitem[{{Rockosi} {et~al.}(2022){Rockosi}, {Lee}, {Morrison}, {Yanny},
  {Johnson}, {Lucatello}, {Sobeck}, {Beers}, {Allende Prieto}, {An}, {Bizyaev},
  {Blanton}, {Casagrande}, {Eisenstein}, {Gould}, {Gunn}, {Harding}, {Ivans},
  {Jacobson}, {Janesh}, {Knapp}, {Kollmeier}, {L{\'e}pine},
  {L{\'o}pez-Corredoira}, {Ma}, {Newberg}, {Pan}, {Prchlik}, {Sayers},
  {Schlesinger}, {Simmerer}, \& {Weinberg}}]{2022ApJS..259...60R}
{Rockosi}, C.~M., {Lee}, Y.~S., {Morrison}, H.~L., {et~al.} 2022, \apjs, 259,
  60

\bibitem[{{Sakari} {et~al.}(2018){Sakari}, {Placco}, {Farrell}, {Roederer},
  {Wallerstein}, {Beers}, {Ezzeddine}, {Frebel}, {Hansen}, {Holmbeck},
  {Sneden}, {Cowan}, {Venn}, {Davis}, {Matijevi{\v{c}}}, {Wyse},
  {Bland-Hawthorn}, {Chiappini}, {Freeman}, {Gibson}, {Grebel}, {Helmi},
  {Kordopatis}, {Kunder}, {Navarro}, {Reid}, {Seabroke}, {Steinmetz}, \&
  {Watson}}]{2018ApJ...868..110S}
{Sakari}, C.~M., {Placco}, V.~M., {Farrell}, E.~M., {et~al.} 2018, \apj, 868,
  110

\bibitem[{{Sanders} \& {Das}(2018)}]{2018MNRAS.481.4093S}
{Sanders}, J.~L., \& {Das}, P. 2018, \mnras, 481, 4093

\bibitem[{{Schlafly} \& {Finkbeiner}(2011)}]{2011ApJ...737..103S}
{Schlafly}, E.~F., \& {Finkbeiner}, D.~P. 2011, \apj, 737, 103

\bibitem[{{Schlegel} {et~al.}(1998){Schlegel}, {Finkbeiner}, \&
  {Davis}}]{1998ApJ...500..525S}
{Schlegel}, D.~J., {Finkbeiner}, D.~P., \& {Davis}, M. 1998, \apj, 500, 525

\bibitem[{{Shank} {et~al.}(2022{\natexlab{a}}){Shank}, {Komater}, {Beers},
  {Placco}, \& {Huang}}]{2022ApJS..261...19S}
{Shank}, D., {Komater}, D., {Beers}, T.~C., {Placco}, V.~M., \& {Huang}, Y.
  2022{\natexlab{a}}, \apjs, 261, 19

\bibitem[{{Shank} {et~al.}(2022{\natexlab{b}}){Shank}, {Beers}, {Placco},
  {Limberg}, {Jaques}, {Yuan}, {Schlaufman}, {Casey}, {Huang}, {Lee},
  {Hattori}, \& {Santucci}}]{2022ApJ...926...26S}
{Shank}, D., {Beers}, T.~C., {Placco}, V.~M., {et~al.} 2022{\natexlab{b}},
  \apj, 926, 26

\bibitem[{{Soubiran} {et~al.}(2016){Soubiran}, {Le Campion}, {Brouillet}, \&
  {Chemin}}]{2016A&A...591A.118S}
{Soubiran}, C., {Le Campion}, J.-F., {Brouillet}, N., \& {Chemin}, L. 2016,
  \aap, 591, A118

\bibitem[{{Starkenburg} {et~al.}(2017){Starkenburg}, {Martin}, {Youakim},
  {Aguado}, {Allende Prieto}, {Arentsen}, {Bernard}, {Bonifacio}, {Caffau},
  {Carlberg}, {C{\^o}t{\'e}}, {Fouesneau}, {Fran{\c{c}}ois}, {Franke},
  {Gonz{\'a}lez Hern{\'a}ndez}, {Gwyn}, {Hill}, {Ibata}, {Jablonka},
  {Longeard}, {McConnachie}, {Navarro}, {S{\'a}nchez-Janssen}, {Tolstoy}, \&
  {Venn}}]{2017MNRAS.471.2587S}
{Starkenburg}, E., {Martin}, N., {Youakim}, K., {et~al.} 2017, \mnras, 471,
  2587

\bibitem[{{Steinmetz} {et~al.}(2006){Steinmetz}, {Zwitter}, {Siebert},
  {Watson}, {Freeman}, {Munari}, {Campbell}, {Williams}, {Seabroke}, {Wyse},
  {Parker}, {Bienaym{\'e}}, {Roeser}, {Gibson}, {Gilmore}, {Grebel}, {Helmi},
  {Navarro}, {Burton}, {Cass}, {Dawe}, {Fiegert}, {Hartley}, {Russell},
  {Saunders}, {Enke}, {Bailin}, {Binney}, {Bland-Hawthorn}, {Boeche}, {Dehnen},
  {Eisenstein}, {Evans}, {Fiorucci}, {Fulbright}, {Gerhard}, {Jauregi}, {Kelz},
  {Mijovi{\'c}}, {Minchev}, {Parmentier}, {Pe{\~n}arrubia}, {Quillen}, {Read},
  {Ruchti}, {Scholz}, {Siviero}, {Smith}, {Sordo}, {Veltz}, {Vidrih}, {von
  Berlepsch}, {Boyle}, \& {Schilbach}}]{2006AJ....132.1645S}
{Steinmetz}, M., {Zwitter}, T., {Siebert}, A., {et~al.} 2006, \aj, 132, 1645

\bibitem[{{Str{\"o}mgren}(1956)}]{1956VA......2.1336S}
{Str{\"o}mgren}, B. 1956, Vistas in Astronomy, 2, 1336

\bibitem[{{Suda} {et~al.}(2008){Suda}, {Katsuta}, {Yamada}, {Suwa}, {Ishizuka},
  {Komiya}, {Sorai}, {Aikawa}, \& {Fujimoto}}]{2008PASJ...60.1159S}
{Suda}, T., {Katsuta}, Y., {Yamada}, S., {et~al.} 2008, \pasj, 60, 1159

\bibitem[{{Sun} {et~al.}(2022){Sun}, {Yuan}, \& {Chen}}]{2022ApJS..260...17S}
{Sun}, Y., {Yuan}, H., \& {Chen}, B. 2022, \apjs, 260, 17

\bibitem[{{Wolf} {et~al.}(2018){Wolf}, {Onken}, {Luvaul}, {Schmidt}, {Bessell},
  {Chang}, {Da Costa}, {Mackey}, {Martin-Jones}, {Murphy}, {Preston}, {Scalzo},
  {Shao}, {Smillie}, {Tisserand}, {White}, \& {Yuan}}]{2018PASA...35...10W}
{Wolf}, C., {Onken}, C.~A., {Luvaul}, L.~C., {et~al.} 2018, \pasa, 35, e010

\bibitem[{{Xue} {et~al.}(2014){Xue}, {Ma}, {Rix}, {Morrison}, {Harding},
  {Beers}, {Ivans}, {Jacobson}, {Johnson}, {Lee}, {Lucatello}, {Rockosi},
  {Sobeck}, {Yanny}, {Zhao}, \& {Allende Prieto}}]{2014ApJ...784..170X}
{Xue}, X.-X., {Ma}, Z., {Rix}, H.-W., {et~al.} 2014, \apj, 784, 170

\bibitem[{{Yanny} {et~al.}(2009){Yanny}, {Rockosi}, {Newberg}, {Knapp},
  {Adelman-McCarthy}, {Alcorn}, {Allam}, {Allende Prieto}, {An}, {Anderson},
  {Anderson}, {Bailer-Jones}, {Bastian}, {Beers}, {Bell}, {Belokurov},
  {Bizyaev}, {Blythe}, {Bochanski}, {Boroski}, {Brinchmann}, {Brinkmann},
  {Brewington}, {Carey}, {Cudworth}, {Evans}, {Evans}, {Gates}, {G{\"a}nsicke},
  {Gillespie}, {Gilmore}, {Nebot Gomez-Moran}, {Grebel}, {Greenwell}, {Gunn},
  {Jordan}, {Jordan}, {Harding}, {Harris}, {Hendry}, {Holder}, {Ivans},
  {Ivezi{\v{c}}}, {Jester}, {Johnson}, {Kent}, {Kleinman}, {Kniazev},
  {Krzesinski}, {Kron}, {Kuropatkin}, {Lebedeva}, {Lee}, {French Leger},
  {L{\'e}pine}, {Levine}, {Lin}, {Long}, {Loomis}, {Lupton}, {Malanushenko},
  {Malanushenko}, {Margon}, {Martinez-Delgado}, {McGehee}, {Monet}, {Morrison},
  {Munn}, {Neilsen}, {Nitta}, {Norris}, {Oravetz}, {Owen}, {Padmanabhan},
  {Pan}, {Peterson}, {Pier}, {Platson}, {Re Fiorentin}, {Richards}, {Rix},
  {Schlegel}, {Schneider}, {Schreiber}, {Schwope}, {Sibley}, {Simmons},
  {Snedden}, {Allyn Smith}, {Stark}, {Stauffer}, {Steinmetz}, {Stoughton},
  {SubbaRao}, {Szalay}, {Szkody}, {Thakar}, {Sivarani}, {Tucker}, {Uomoto},
  {Vanden Berk}, {Vidrih}, {Wadadekar}, {Watters}, {Wilhelm}, {Wyse}, {Yarger},
  \& {Zucker}}]{2009AJ....137.4377Y}
{Yanny}, B., {Rockosi}, C., {Newberg}, H.~J., {et~al.} 2009, \aj, 137, 4377

\bibitem[{{York} {et~al.}(2000){York}, {Adelman}, {Anderson}, {Anderson},
  {Annis}, {Bahcall}, {Bakken}, {Barkhouser}, {Bastian}, {Berman}, {Boroski},
  {Bracker}, {Briegel}, {Briggs}, {Brinkmann}, {Brunner}, {Burles}, {Carey},
  {Carr}, {Castander}, {Chen}, {Colestock}, {Connolly}, {Crocker}, {Csabai},
  {Czarapata}, {Davis}, {Doi}, {Dombeck}, {Eisenstein}, {Ellman}, {Elms},
  {Evans}, {Fan}, {Federwitz}, {Fiscelli}, {Friedman}, {Frieman}, {Fukugita},
  {Gillespie}, {Gunn}, {Gurbani}, {de Haas}, {Haldeman}, {Harris}, {Hayes},
  {Heckman}, {Hennessy}, {Hindsley}, {Holm}, {Holmgren}, {Huang}, {Hull},
  {Husby}, {Ichikawa}, {Ichikawa}, {Ivezi{\'c}}, {Kent}, {Kim}, {Kinney},
  {Klaene}, {Kleinman}, {Kleinman}, {Knapp}, {Korienek}, {Kron}, {Kunszt},
  {Lamb}, {Lee}, {Leger}, {Limmongkol}, {Lindenmeyer}, {Long}, {Loomis},
  {Loveday}, {Lucinio}, {Lupton}, {MacKinnon}, {Mannery}, {Mantsch}, {Margon},
  {McGehee}, {McKay}, {Meiksin}, {Merelli}, {Monet}, {Munn}, {Narayanan},
  {Nash}, {Neilsen}, {Neswold}, {Newberg}, {Nichol}, {Nicinski}, {Nonino},
  {Okada}, {Okamura}, {Ostriker}, {Owen}, {Pauls}, {Peoples}, {Peterson},
  {Petravick}, {Pier}, {Pope}, {Pordes}, {Prosapio}, {Rechenmacher}, {Quinn},
  {Richards}, {Richmond}, {Rivetta}, {Rockosi}, {Ruthmansdorfer}, {Sandford},
  {Schlegel}, {Schneider}, {Sekiguchi}, {Sergey}, {Shimasaku}, {Siegmund},
  {Smee}, {Smith}, {Snedden}, {Stone}, {Stoughton}, {Strauss}, {Stubbs},
  {SubbaRao}, {Szalay}, {Szapudi}, {Szokoly}, {Thakar}, {Tremonti}, {Tucker},
  {Uomoto}, {Vanden Berk}, {Vogeley}, {Waddell}, {Wang}, {Watanabe},
  {Weinberg}, {Yanny}, {Yasuda}, \& {SDSS Collaboration}}]{2000AJ....120.1579Y}
{York}, D.~G., {Adelman}, J., {Anderson}, John~E., J., {et~al.} 2000, \aj, 120,
  1579

\bibitem[{{Yuan} {et~al.}(2013){Yuan}, {Liu}, \& {Xiang}}]{2013MNRAS.430.2188Y}
{Yuan}, H.~B., {Liu}, X.~W., \& {Xiang}, M.~S. 2013, \mnras, 430, 2188

\bibitem[{{Yuan} {et~al.}(2022){Yuan}, {Malhan}, {Sestito}, {Ibata}, {Martin},
  {Chang}, {Li}, {Caffau}, {Bonifacio}, {Bellazzini}, {Huang}, {Voggel},
  {Longeard}, {Arentsen}, {Doliva-Dolinsky}, {Navarro}, {Famaey},
  {Starkenburg}, \& {Aguado}}]{2022ApJ...930..103Y}
{Yuan}, Z., {Malhan}, K., {Sestito}, F., {et~al.} 2022, \apj, 930, 103

\bibitem[{{Zepeda} {et~al.}(2022){Zepeda}, {Rasmussen}, {Beers}, {Placco},
  {Huang}, \& {Depagne}}]{2022ApJ...927...13Z}
{Zepeda}, J., {Rasmussen}, K.~C., {Beers}, T.~C., {et~al.} 2022, \apj, 927, 13

\bibitem[{{Zhang} \& {Yuan}(2023)}]{2023ApJS..264...14Z}
{Zhang}, R., \& {Yuan}, H. 2023, \apjs, 264, 14

\bibitem[{{Zheng} {et~al.}(2018){Zheng}, {Zhao}, {Wang}, {Fan}, {Tan}, {Li}, \&
  {Zuo}}]{2018RAA....18..147Z}
{Zheng}, J., {Zhao}, G., {Wang}, W., {et~al.} 2018, Research in Astronomy and
  Astrophysics, 18, 147

\bibitem[{{Zhou} {et~al.}(2016){Zhou}, {Fan}, {Fan}, {He}, {Jiang}, {Jiang},
  {Jing}, {Lesser}, {Ma}, {Nie}, {Shen}, {Wang}, {Wu}, {Zhang}, {Zhou}, \&
  {Zou}}]{2016RAA....16...69Z}
{Zhou}, X., {Fan}, X.-H., {Fan}, Z., {et~al.} 2016, Research in Astronomy and
  Astrophysics, 16, 69

\end{thebibliography}
\end{document}